\documentclass[11pt, a4paper]{article}
\pdfoutput=1
\usepackage{graphicx}
\usepackage{jcappub}

\usepackage{dcolumn}
\usepackage{bm}
\usepackage{subfigure}
\usepackage{wasysym}
\usepackage{hyperref}
\usepackage{comment}

\newcommand{\be}{\begin{equation}}
\newcommand{\ee}{\end{equation}}

\renewcommand{\vec}{\bm}

\newcommand{\bdelta}{b_\delta}

\newcommand{\vpar}{v_\parallel}

\newcommand{\lya}{Ly$\alpha$}

\newcommand{\lyaf}{Ly$\alpha$ forest}

\newcommand{\bF}{\bar{F}}

\begin{document}

\title{Towards physics responsible for large-scale Lyman-\boldmath$\alpha$
  forest bias parameters}

\author{Agnieszka M. Cieplak,}
\author{An\v{z}e Slosar}
\affiliation{Brookhaven National Laboratory, \\
             Bldg 510A, Upton NY 11973, USA}
\emailAdd{acieplak@bnl.gov}
\emailAdd{anze@bnl.gov}

\date{\today}
\newcommand{\bieta}{b_\eta}
\newcommand{\btau}{\bar{\tau}}
\renewcommand{\lyaf}{Lyman-$\alpha$ forest}
\renewcommand{\lya}{Lyman-$\alpha$}

\abstract{
  Using a series of carefully constructed numerical experiments based
  on hydrodynamic cosmological SPH simulations, we attempt to build an intuition for the
  relevant physics behind the large scale density ($\bdelta$) and
  velocity gradient ($\bieta$) biases of the Lyman-$\alpha$
  forest. Starting with the fluctuating Gunn-Peterson approximation
  applied to the smoothed total density field in real-space, and
  progressing through redshift-space with no thermal broadening,
  redshift-space with thermal broadening and hydrodynamicaly
  simulated baryon fields, we investigate how approximations found in
  the literature fare. We find that Seljak's 2012 analytical formulae for these bias parameters 
  work surprisingly well in the limit of no thermal broadening and linear
  redshift-space distortions. We also show that his $\bieta$ formula
  is exact in the limit of no thermal broadening. Since introduction
  of thermal broadening significantly affects its value, we speculate
  that a combination of large-scale measurements of $\bieta$ and the small
  scale flux PDF might be a sensitive probe of the thermal state of the
  IGM.  We find that large-scale biases derived from the smoothed
  total matter field are within 10-20\% to those based on
  hydrodynamical quantities, in line with other measurements in the
  literature.}
  
 
 \arxivnumber{1509.07875}

\maketitle

\section{Introduction}
Measuring the growth of large scale structure has become a powerful
probe of testing cosmological parameters. The \lyaf\ has emerged as
one of the primary tracers of this large scale structure at
intermediate redshifts. Seen as absorption features in quasar spectra,
the forest probes the distribution of neutral hydrogen between the
observer and the quasar. The \lya\ absorption of neutral hydrogen has
a particularly large cross section, and as the emission continuum of a
quasar becomes redshifted to this transition wavelength, it will be
absorbed even with small fractions of neutral hydrogen present. This
method is therefore sensitive to low gas densities, serving as a
tracer of the large scale matter distribution between the redshifts of
1.7 (below which the gas becomes fully ionised while at the same time
the UV light continuum becomes redshifted to wavelengths absorbed by
the atmosphere) and redshifts 4-5 (above which the forest becomes
opaque). At small scales, the \lyaf\ has given us unique constraints
on, for example, warm dark matter models \cite{2013PhRvD..88d3502V,
  2006PhRvL..97s1303S} and primordial black hole dark matter
\cite{2003ApJ...594L..71A}, whereas on intermediate scales it has
probed limits on the sum of the neutrino masses
\cite{2015JCAP...02..045P, LESGOURGUES2015, 2006JCAP...10..014S},
the running of the spectral index
\cite{2006JCAP...10..014S}, or dark matter - baryon scattering \cite{2014PhRvD..89b3519D}.
Recently, at large scales, it has provided
a detection of the baryon acoustic oscillations at intermediate
redshifts \cite{2013A&A...552A..96B, 2013JCAP...04..026S,
  2015A&A...574A..59D}. With the completion of the BOSS survey
\cite{2013AJ....145...10D}, and the start of eBOSS \cite{eBOSS}, as
well as the planned DESI survey \cite{2014JCAP...05..023F}, future
\lyaf\ measurements will be covering larger scales with
greater precision. It is therefore crucial that we also improve our
understanding of the biases involved between what is observed and the
underlying matter distribution.

The relationship between the measured \lyaf\ flux and the underlying
matter is highly nonlinear, but the physics is thought to be
well-understood. Underlying neutral hydrogen gas is related to this
measured flux via the optical depth, $\tau=-\mathrm{ln}F$, which is
proportional to the \lya\ absorption cross section and the number of
neutral hydrogen atoms along the line of sight (los). If
photoionization equilibrium at these redshifts is assumed, the amount
of neutral hydrogen is related in a temperature-dependent way to the
total number of hydrogen atoms, which in turn traces the underlying
matter density. Additionally, assuming adiabatic expansion, which
produces a tight temperature-density relation, $\gamma-1=d \mathrm{ln}
\rho/d \mathrm{ln} T$ \cite{1997MNRAS.292...27H}, the optical depth can be
modeled as $\tau = A (1+\delta)^\alpha$, where $\delta$ is the gas
overdensity, $A$ is a constant which depends on the photoionization
rate, the temperature of the gas, and redshift, while $\alpha =
2-0.7(\gamma-1)$. The gas overdensity traces the total matter
overdensity down to the Jeans scale, where gas pressure supports
against gravitational collapse.

This Fluctuating Gunn Peterson Approximation (FGPA) \cite{1998ApJ...495...44C}, although
it ignores shock heated gas and thermal broadening, has been
surprisingly successful in explaining statistical properties of the
\lyaf\ observations. Because it is a local transformation, the flux
fluctuations, defined as $\delta_F = F/\bar{F} -1$ (with $\bar{F}$
being the mean absorbed flux) will trace the
dynamically dominant fields on large scales \cite{2003ApJ...585...34M,2015arXiv150604519A}:
\begin{equation}
  \delta_F = \bdelta \delta + \bieta \eta + b_\Gamma \delta_\Gamma + \epsilon.
\end{equation}
This equation is valid in Fourier-space in the limit $k\rightarrow 0$
or on a sufficiently smoothed real-space field (i.e where fluctuations
are smoothed so that the Taylor expansion in $\delta$ is valid). The
quantities $\delta$, $\delta_\eta$ and $\delta_\Gamma$ are relative
fluctuations in the total density, mass-weighted velocity gradient
$\eta = -H^{-1} dv_{\vert\vert}/dr_{\vert\vert}$ and photo-ionization
fluctuations rate $\Gamma$. 
The bias parameters $\bdelta$, $\bieta$
and $b_\Gamma$ encode properties of the small-scale physics that
determines their values on large-scales. In this paper we will use
these symbols to denote absorbed flux fluctuation ($\delta_F$) biases, but biases for other
fields will denote tracer field, e.g. $b_{\tau,\delta}$ for the density
bias of the $\delta_\tau$ field (all tracer fields will be normalized to the mean).
Finally, $\epsilon$ describes the scatter
between the tracer field $\delta_F$ and the prediction from source fields. It
is expected to have a white power $P_N$ on large scales. We will not
consider photoionization fluctuations in this paper, but see
e.g.\cite{2014PhRvD..89h3010P, 2014arXiv1404.7425G} and references there-in.

Ignoring the $\Gamma$ field, it then follows that the three-dimensional power-spectrum of flux on
large scales  is given by
\begin{equation}
P_{\delta_F}(\vec{k})=\bdelta^2 \left(1+\beta \mu^2\right)^2
P_\delta(k) + P_N
\label{eq: powerspectrum}
\end{equation}
where $\beta=f b_\eta/b_{\delta}$, with $f$ being the logarithmic
growth rate, $\mu$ is the cosine of the angle between the los and the
vector $\vec{k}$, and $P_\delta$ is the total matter power
spectrum. This is equivalent to the famous Kaiser \cite{1987MNRAS.227....1K}
redshift-space distortion (RSD) formula for galaxies, but note that in our
case $\beta$ is not uniquely determined by $\bdelta$, as we will
discuss later in the paper. The noise power $P_N$ is expected to be
small and is often ignored.

Parameters $\bdelta$ and $\bieta$ (or equivalently $\beta$) can be
thought of as purely phenomenological parameters to be marginalised
over as is done in e.g. BAO analyses. On the other hand they encode
the same small scale physics that is used when predicting small scale 1D
power spectra that has been used to constrain cosmology. Therefore, 
with a goal to eventually be able to perform
a unified analysis of correlations in the \lyaf, from smallest to the
largest available scales, rather then artificially splitting the
problem into a small-scale 1D power spectrum and a large-scale correlation
function, the underlying physics that determines the
values of these parameters has to be well-understood.

This paper attempts to bring some understanding to what determines the
numerical value of these large-scale biases and how well these can be
described using analytical predictions. The main focus in this study is not
on using these predictions for precision cosmology, but primarily to first 
understand how accurate these predictions are or when do they fail in 
describing results of $N$-body simulations. This expands and
complements previous work trying to understand the large scale
fluctuations in the \lyaf\ \cite{2002MNRAS.329..848V, 2015JCAP...04..009W, 2003ApJ...585...34M,2009JCAP...10..019S,
2011MNRAS.413.1717B,2011MNRAS.415..977M,2015arXiv150604519A}.

\section{Theory}

The basic starting point for derivation of biases is the notion of
peak-background split (PBS)\footnote{The etymology of the wording
  ``peak-background'' split comes from thinking about the evolution of
  dark matter halos as separate from the evolution of the large-scale
  overdensity -- for all practical purposes the halo cannot see the
  difference between a different background cosmology and a sufficiently
  large-wavelength mode. The same idea often goes under the name
  ``separate universe'' method. Of course, in our case, there are no
  halos, but we resign to use this widely adopted phrase to refer to
  the fact that small-scale non-linear evolution cannot distinguish
  between a large-scale linear mode and a modified background.},
namely that a large over-dense region of the universe is equivalent to a
slightly over-dense universe and that a large region of universe with
non-zero $\eta$ is equivalent to an anisotropic universe with a
velocity gradient. This picture is exact in the $k\rightarrow 0$ limit
and gives the following simple prescription for the calculation of bias
parameters\cite{2003ApJ...585...34M,2012JCAP...03..004S,2015arXiv150604519A}:
\begin{eqnarray}
 b_{\delta} &=& \frac{1}{\bF} \frac{d\bF}{d\delta}  \vert_{\eta=0} \label{eq:pbsplit1}\\
 b_\eta &=& \frac{1}{\bF} \frac{d\bF}{d\eta}  \vert_{\delta=0} \label{eq:pbsplit2}
\end{eqnarray}
These formulae can be used both in theoretical modeling and in
simulations.

This approach was used by \cite{2012JCAP...03..004S} to make a first analytical
attempt at understanding the large scale bias and redshift space
distortion parameter. By using the second order perturbation theory
gravitational coupling between short and long wavelength modes, 
 analytical formulae can be obtained that link $\bdelta$ and $\bieta$
to the observed flux probability distribution function (PDF). 

Approximations used in that paper were 

\begin{itemize}
\item The second order standard perturbation theory gives sufficiently accurate
  predictions bias of density field to $n$-th power
\item The flux field is fully determined by the FGPA formula
  \begin{equation}
    F=\exp\left[(-A(1+\delta)^\alpha\right]
    \label{eq:FGPA}
  \end{equation}
\item No thermal broadening of the lines
\end{itemize}
Somewhat surprisingly,  we will later discover that the inaccuracy 
of these equations is mostly due to thermal broadening and the
non-linearity of the velocity field.

\subsection{Density bias}

We will proceed to review the derivation of bias parameters. Our
treatment is somewhat different from that of
\cite{2012JCAP...03..004S} in that it makes the peak-background
argument more obvious, but we arrive at identical results.  In the
presence of a long wavelength density mode, $\delta_l$, in an Einstein-deSitter universe,
the short wavelength density modes, $\delta_s$, in second order perturbation theory become
\begin{equation}
\delta_s^{\mathrm{pert}} = (1+\nu_2 \delta_l)\delta_s,
\label{eq:2ndorder}
\end{equation}
where $\nu_2=34/21$, is the angular average of the second order
perturbation theory kernel $F_2(\vec{k_1},\vec{k_2})$. This relation
does not require that $\delta_s$ modes are linear, it simply provides
an approximate mapping between a set of non-linear modes in a patch of
universe as a function of large-scale over-density.

Therefore, the total density transforms as
\begin{equation}
\delta = \delta_s^{\mathrm{pert}}+\delta_l = (1+\nu_2 \delta_l)\delta_s+\delta_l.
\label{eq:1}
\end{equation}
The mean for a quantity $X$ in the presence of large-scale mode $\delta_l$
are thus given by
\begin{equation}
  \bar{X} = \int_{-\infty}^\infty 
  X(\delta=\delta(1+\nu_2\delta_l)+\delta_l) p(\delta) d\delta,
\end{equation}
where $p(\delta)$ is the probability distribution function for
$\delta$. Using Equation \eqref{eq:pbsplit1}, this allows us to
calculate bias for any quantity which is uniquely determined by
$\delta$ and connect it to its PDF.  Applying this to matter fields to
the $n$-th power ($\delta^n$) (and normalizing by the mean of this tracer field), we find
\begin{equation}
b_{\delta^n,\delta} = n \nu_2  + n \langle \delta_s^{n-1} \rangle /\langle \delta_s^n \rangle.
\label{eq:moments}
\end{equation}
The large scale bias for the \lyaf\  optical depth with
respect to the underlying density is given by
\begin{equation}
 b_{\tau,\delta} = \btau^{-1} \frac{\partial \btau}{\partial \delta_l} =
 \btau^{-1} \nu_2 \left\langle \delta \frac{d\btau}{d\delta}  \right\rangle + \btau^{-1} \left\langle \frac{d \btau}{d \delta} \right\rangle.
\end{equation}
For the simple isothermal density-temperature relation, where $\alpha
= 2$, this gives
\begin{equation}
b_{\tau,\delta}=\frac{2(1+\nu_2 \sigma^2_J)}{1+\sigma^2_J},
\end{equation}
where $\sigma_J$ is the rms density field smoothed on a Jeans
scale. 

The bias for optical depth is a theoretically interesting quantity, but it
is not what we need. Ignoring redshift space distortions in the optical depth
for now, the flux bias relating the flux to the underlying density
field, can then be calculated as:
\begin{equation}
b_\delta = \frac{1}{\bF} \frac{\partial \bF[\tau({\delta)}]}{\partial \delta_l} = \frac{1}{\bF} \left[\nu_2 \left \langle \delta \frac{dF}{d\delta} \right \rangle + \left \langle \frac{dF}{d \delta} \right \rangle \right].
\end{equation}

Using Equation \ref{eq:FGPA}, the above bias
can be expressed in terms of the constants $A$ and $\alpha$:

\begin{equation}
b_\delta =  \bar{F}^{-1} [\alpha \langle F \mathrm{ln} F \rangle +
  \alpha (\nu_2-1) \left \langle F \mathrm{ln} F [1-(-\mathrm{ln}
    F/A)^{-\alpha^{-1}}] \right \rangle ].
\end{equation}

Remarkably, therefore, this theoretical bias could be calculated from
just the constants $A$ and $\alpha$, which could be derived from the
temperature-density scatter plots from simulations, and the observed flux PDF
in the data.

\subsection{Velocity gradient bias}
For the velocity gradient bias, consider the effect of $\eta$ on a
large patch of the universe. The redshift-space coordinate $s=r + H^{-1}\vpar$
transforms as
\begin{equation}
  s \rightarrow s - r\eta.
\label{eq:etap}
\end{equation}
Action of the $\eta$ field is thus equivalent to stretching the
redshift-space in the radial direction. Since stretching cannot change
the total number of hydrogen atoms, the mean value of $\tau$ must
change by $(1-\eta)^{-1}$ giving $b_{\tau,\eta}=1$. The same argument
holds for all other fields that are conserved on transformation from
real to redshift space (e.g. number of galaxies\footnote{As an
  interesting aside: One could imagine that the mean value of a field
  would respond non-trivially to a global anisotropic expansion even
  in real-space, which would then lead to $\bieta\neq 1$ even for
  tracers that are conserved. But one can only imagine it. If this was
  the case, then, by symmetry the bias factors for $\eta_x=dv_x/dx$
  and $\eta_y=dv_y/dy$ should be the same (assuming $x$ and $y$
  coordinates are transversal to the radial $z$ coordinate). Since
  $\eta_x+\eta_y+\eta_z \propto \delta$ on the large, linear scales, it
  is clear that there can be no change in the mean field on average due to a
  change in $\eta$ at fixed $\delta=0$. There will be, however,
  second-order effects.}).  In the limit of no
thermal broadening, the $\tau$ simply responds as pure rescaling
\begin{equation}
  \tau(r) \rightarrow \frac{\tau\left(r(1-\eta)^{-1}\right)}{1-\eta} = \tau\left(r(1-\eta)^{-1}\right)(1+\eta) +O(\eta^2)
\label{eq:taupb}
\end{equation}
Using Equation \ref{eq:pbsplit2}, 
\begin{equation}
  \bF = \int_0^1 p(F) F dF
\end{equation}
and $\tau=-\ln F$, one can derive
\begin{equation}
b_\eta = \left \langle F \mathrm{ln} F \right \rangle.
\end{equation}

This equation was derived by Seljak by starting with the relation
$\tau(\vec{s}) = \tau(\vec{r})(1+f \mu^2 \delta)$ and then Taylor
expanding the expression for $F$. The above shows that it is in fact
much more general and gives the correct answer on large scales even if
the distortions are not Kaiser-like in $\tau$ on intermediate scales,
as long as velocity dispersion is negligible.  However, as we will see
later, the presence of thermal broadening is significant and breaks
this formula, since the broadening kernel does not stretch with the
$(1-\eta)$ factor.

In this paper we use SPH simulations to test these analytical
formulae. We start with the smoothed total matter field part of these simulations and apply the FPGA
equation exactly in real space to see how well they fare in this
idealised toy-case. We then proceed to progressively more realistic
cases, ending with the analysis of the full hydrodynamic part of the simulation.

\section{Description of Numerical Methods and Simulations Used}

The simulations used for this comparison study are Gadget-3 \cite{Springel2005} hydrodynamic 
simulations with a box size of $L=40$ $\mathrm{Mpc/h}$ with $N=2 \times 1024^3$
for the number of gas and dark matter particles. They are evolved with a Haardt and Madau
UV background \cite{1996ApJ...461...20H} and the simple QUICKLYA option for star formation with no feedback, where any SPH particle that reaches a maximum density is converted into a star particle. The standard critical over-density, as used in these simulations, is 1000 times the mean. The SPH smoothing length is such that the particle's density is estimated by smoothing over 32 neighboring particles. The
fiducial simulation uses the WMAP7 cosmology.

In the following, we start with simply analyzing the density field from these hydrodynamic simulations, using the total matter overdensity output $\delta$ from the simulation at each point to transform to a flux field via the FGPA in Equation 2.3. We progressively add the redshift space distortions (RSD), and thermal broadening by hand to this flux field as described below in Section 4. Before applying the FGPA to the density field, we smooth it using a Gaussian filter
$\delta(k) \to \delta(k) \mathrm{exp}(-k^2 R^2/2)$ to account for
Jeans smoothing present in baryon fields. We always measure our
quantities at different levels of smoothing, including unrealistically
large smoothing factors, since analytical expressions are expected to
perform better at more aggressive levels of smoothing.
Unless stated otherwise, we assume $A=0.3 ((1+z)/(1+2.4))^{4.5}$ and
$\alpha=1.6$ as in \cite{2012JCAP...03..004S}.
Finally, it is only in Section 6, that we analyze the flux field from the hydrodynamic part of these same simulations, in order to compare it to the one generated from this FGPA prescription.

We calculate true bias parameters using two methods, which we describe
in the following two sub-sections.

\subsection{Peak-background split applied to simulations}
This amounts to numerically evaluating expressions in Equations
\ref{eq:pbsplit1} and \ref{eq:pbsplit2} and has been used to measure the bias parameter by
McDonald well over a decade ago \cite{2003ApJ...585...34M}.

\begin{table}
\centering
\begin{tabular}{l | | c | c | c}
\hline
\hline
$\delta(z=2.5)$ & -0.015 & 0.0 & 0.015\\
$\Omega_m (z=0)$ & 0.260 & 0.275 & 0.290\\
$\Omega_k (z=0)$ & 0.024 & 0.0 & -0.024\\
$\Omega_\Lambda (z=0)$ & 0.715 & 0.725 & 0.735\\
$h0$ & 0.707 & 0.702 & 0.697\\
$\sigma_8$ & 0.795 & 0.816 & 0.836\\
\hline
\hline
\end{tabular}
\caption{Parameters of peak-background split simulations. \label{tab:pbsims}}
\end{table}

For density simulations, this entails running two additional
simulations, acting as over- and underdense counterparts to the
fiducial simulation. In order to achieve this, the starting
cosmological parameters for these simulations have to be changed. The
main requirement is for the density in the perturbed universe to
satisfy $\rho'(t) = \rho(t) (1+\delta_l(t))$, where $\delta_l$
represents the matter overdensity with respect to the background
universe (which can be thought of as the long wavelength mode
overdensity in this region of the universe). From this relation, the
cosmologies for the perturbed simulations can be derived. Requiring a
$\delta_l$ value of $+/- 0.015$ at $z=2.5$ (the central redshift of
interest for the \lyaf\ ), the perturbed cosmological
parameters in Appendix \ref{app:pb} can be evaluated at $z=0$ to give
the values in Table \ref{tab:pbsims}.  All three simulations listed in
Table \ref{tab:pbsims} are evolved from $z_i=159$ with outputs at $z=$ 3, 2.8, 2.75,
2.6, 2.5, 2.4, 2.25, 2.2, and 2. Here we focus our analysis on
redshifts 3, 2.5, and 2. Since the evolution of redshift with time is
different in the perturbed universe we rescale the density field of
the perturbed simulation from the nearest redshift output $z$ to match
the required $z'$ as seen in Equation \ref{eq:pbredshift} using the growth
factor. (The matching $z'$ for example to $z=2.5$ is $z'=2.47$ for
$\delta_l=+0.015$). We check that this scaling matches the results
achieved by interpolation methods to better than $0.01\%$.

After the perturbed simulation density fields are rescaled to the
correct redshift, the over-densities are further rescaled to be
expressed with respect to the the mean of the unperturbed simulation
by $\delta^{\mathrm{pert}} \to \delta^{\mathrm{pert}}
(1+\delta_l)+\delta_l$. This $\delta^{\mathrm{pert}} $ is then used to
calculate the mean flux in the FGPA approximation when computing the
numerical derivative in Equations \ref{eq:pbsplit1} and
\ref{eq:pbsplit2} for the peak background split method analysis.

For the velocity gradient bias, the calculation is less painful. The
prescription in Equation \ref{eq:etap} postulates that we need to
imagine our simulation box is now of the size $L^2\times L(1-\eta_l)$,
where we set $\eta_l=\pm 0.01$ to be the long-scale $\eta$ we differentiate
with respect to. This means that before calculating the mean
quantities, we need to i) divide all $\tau$'s by $(1-\eta_l)$ and ii)
divide the thermal broadening parameter by $(1-\eta_l)$. These are
precisely the prescriptions that McDonald gives in his 2003 paper \cite{2003ApJ...585...34M}.
 
\subsection{Direct mode-by-mode measurement}

As a cross-check, we also calculate the bias parameters using the
standard way by comparing the cross matter-trace power spectrum with
the matter power spectrum. We formalize this by writing the likelihood
assuming scatter $\epsilon$ to be normally distributed with variance
$P_N$, i.e.
\begin{equation}
  \log L  = {\rm const.} - \frac{N}{2}  \log P_N - \sum_{{\rm modes\
      i}}  \left(\frac{(\delta_{F,i}-\delta_F(\delta,\eta,\bdelta, \bieta))^2}{2P_N} \right),
\end{equation}
where $\delta_{F,i}$ are the actual measured flux modes and
$\delta_F(\delta,\eta,\bdelta, \bieta)=\bdelta \delta + \bieta \eta$ is
the corresponding prediction from actual total fluctuation modes
$\delta_i$ and bias parameters.
We maximize this likelihood over free parameters $\bdelta$, $\beta$,
$P_N$ and their $k^2$ and $k_\parallel$ corrections (6 parameters in
total) and use the second derivative matrix to derive the marginalised
error assuming Gaussianity.

By analytically maximizing the likelihood, one can show that this
measurement is essentially the same as measuring the bias from the
ratio of the cross-power to the dark matter auto-power
spectrum. Measuring bias from the cross-power spectrum rather than the
flux auto-power spectrum avoids biases due to the presence of the noise
term $P_N$, but this is likely a very small effect. The main advantage
of going through likelihood is that this method correctly propagates
the scatter between the modes into an uncertainty in the bias
determination.

We have tested this method extensively on toy problems. Nevertheless,
to our great consternation, we note that while results are in general
in good agreement with those from the peak-background split, there are
cases where they disagree significantly. We believe that these are due
to two kinds of effects. First, our box size of $40$ Mpc/h is limited
and we do not correctly account for the effect of large-scale
modes. This plagues the PB split method as well, but there at least
the effect of $k=0$ mode is propagated correctly. In fact the
fundamental mode, $k=2\pi/(40\ \mathrm{Mpc/h})$, is on the verge of being significantly non-linear at
$z=2.5$.  Second, although we correctly account for the $k^2$
dependence of bias parameters, we do not fit for other bias parameters
that surely exist at this level, most importantly $b_2$ and $b_{s^2}$
\cite{2009JCAP...08..020M} along with similar parameters for the
$\eta$ field. 

In \cite{2015arXiv150604519A}, the authors take a somewhat different
approach towards fitting bias parameters and fit the power spectrum
measurement with an attempted guess at the shape of non-linear
corrections to the $k\rightarrow 0$ limit. Despite having a larger box
they similarly find that solutions are sensitive to details of the
fitting formulae. 

\section{Results}

We now show our results through a series of plots that use the same
symbol coding. Results from analytical predictions are plotted as 
lines, peak-background split results as dots and mode measurements
as triangles with error-bars. These are usually plotted as a function
of smoothing scale, unless noted otherwise.

\subsection{Moments of the density field}

\begin{figure*}
\begin{tabular}{cc}
   \includegraphics[width=0.5\linewidth]{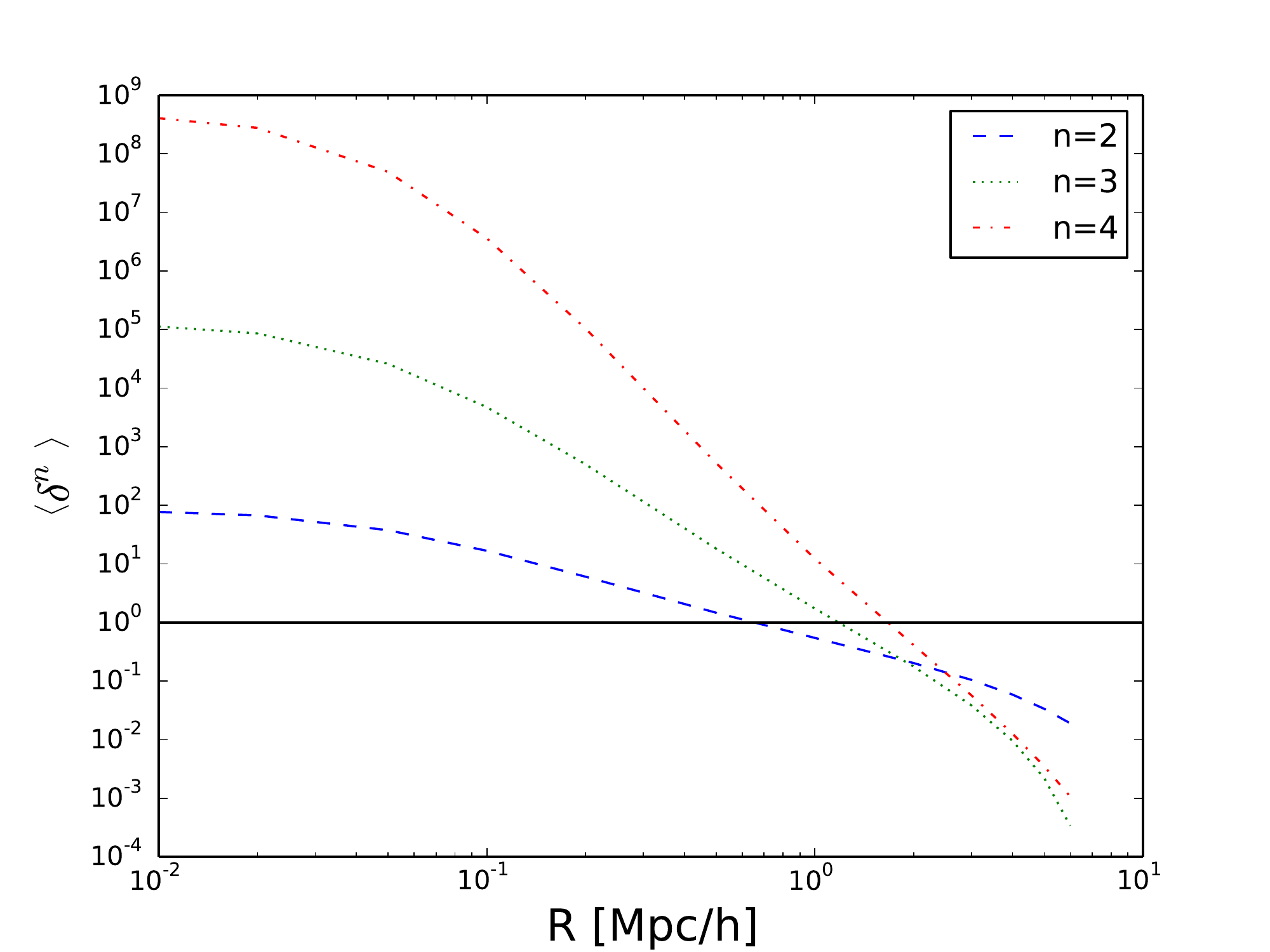} &
  \includegraphics[width=0.5\linewidth]{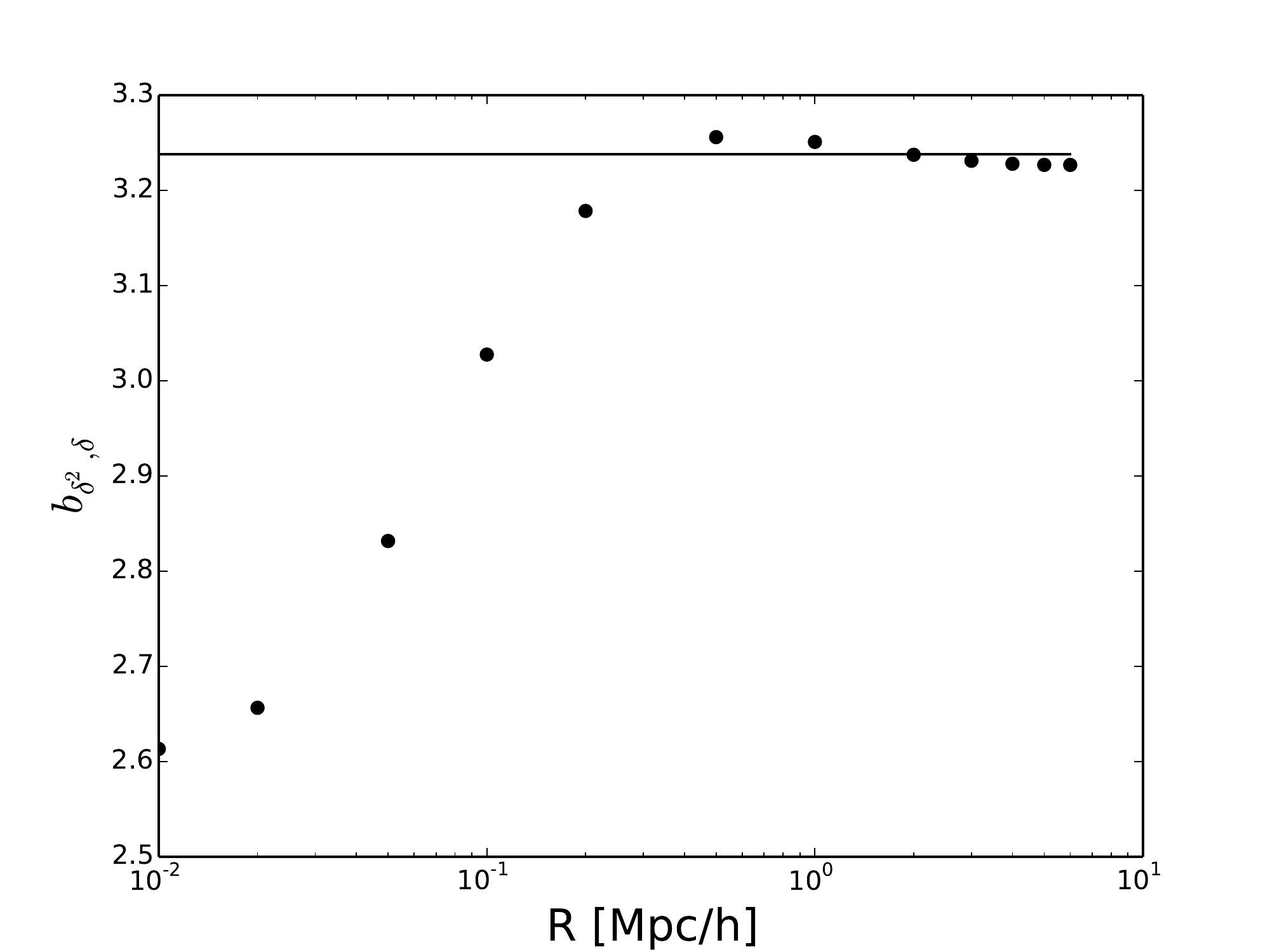} \\
  \includegraphics[width=0.5\linewidth]{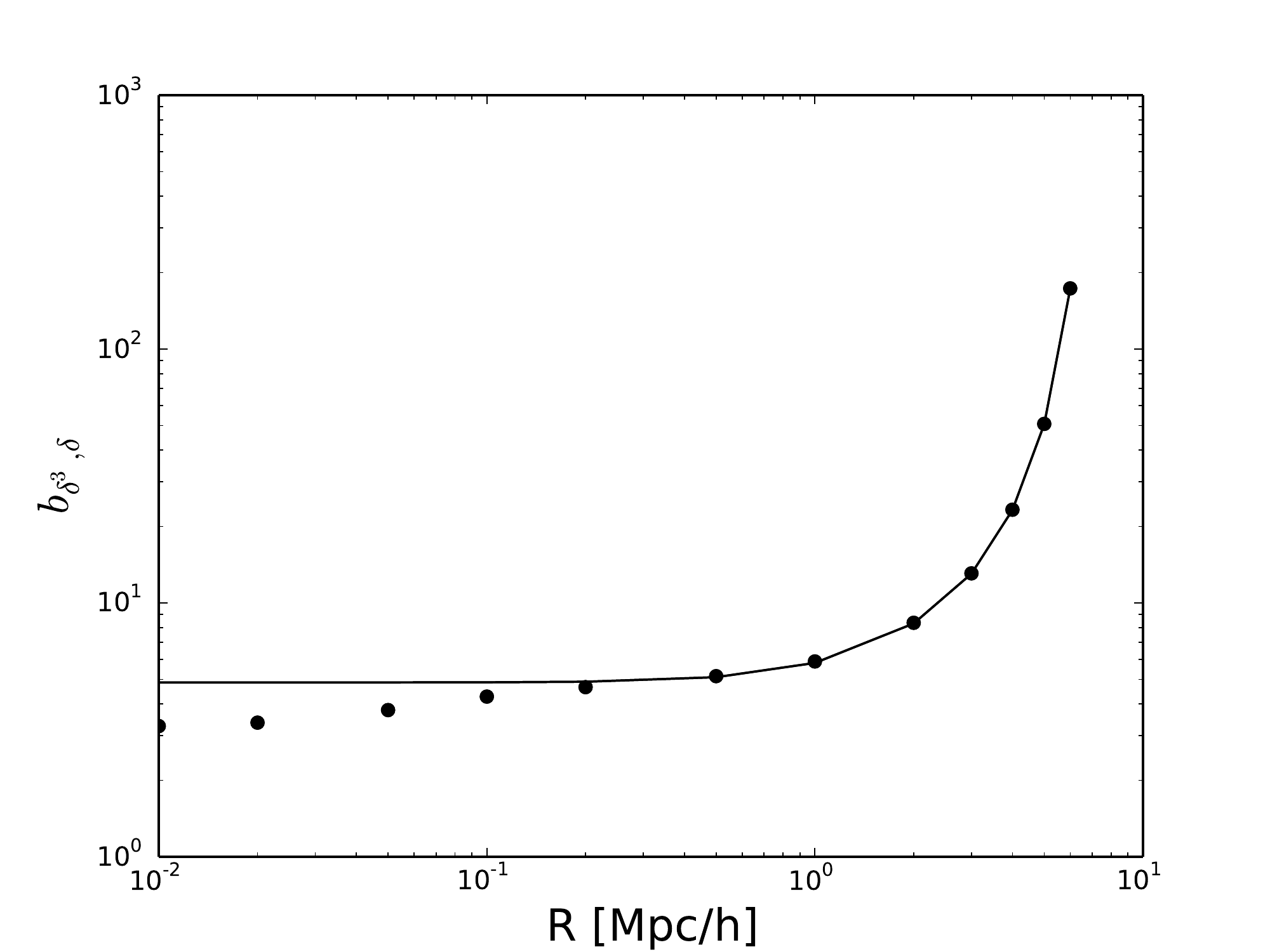} &
  \includegraphics[width=0.5\linewidth]{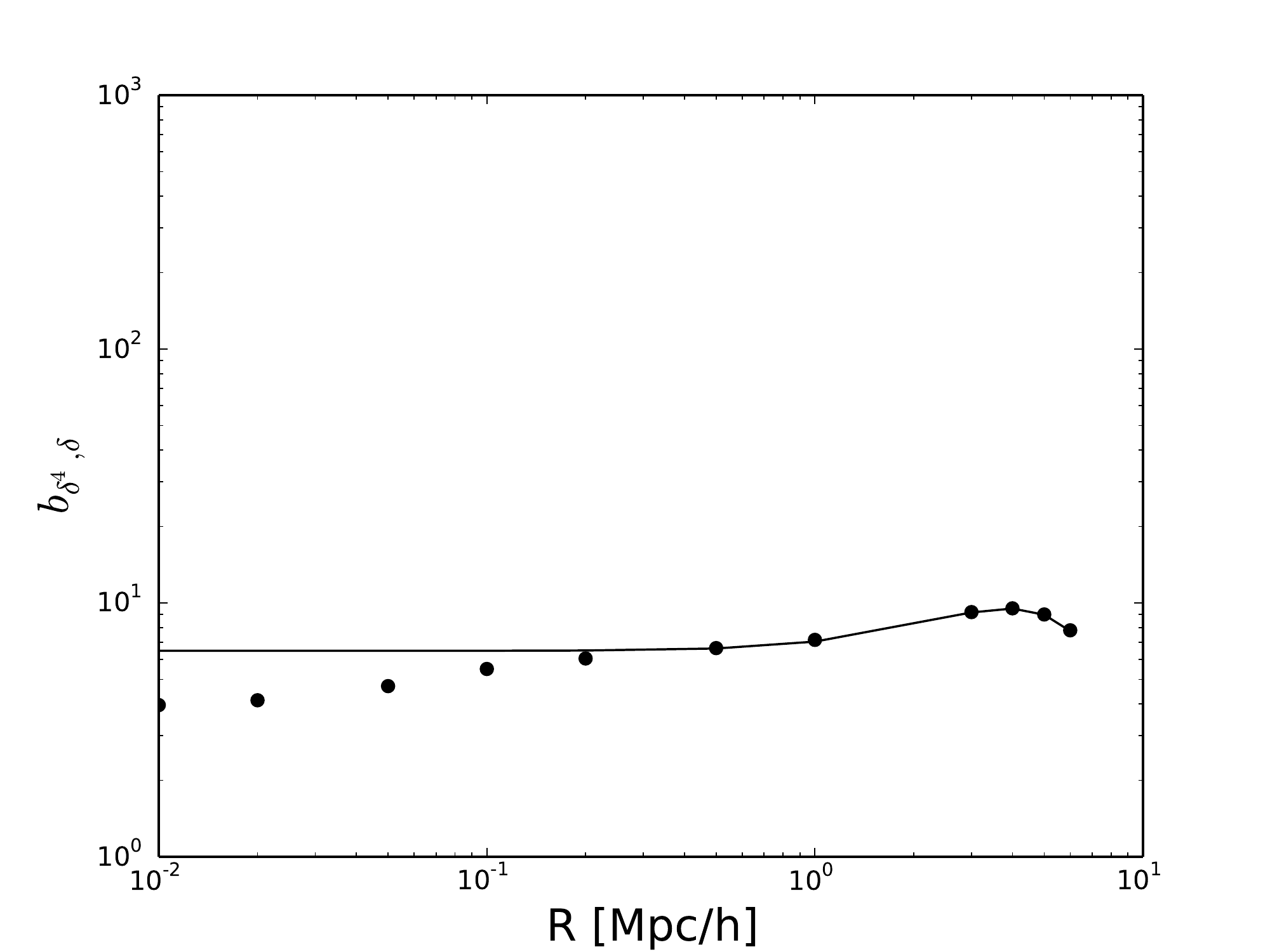} \\
\end{tabular}
	\caption{Upper left panel shows the mean value of moments of
          the matter field $\delta^n$ at redshift 2.5 as a function of smoothing scale.
          Other panels show their bias factors: from peak-background split (filled dots)
          and the analytical prediction (black line).
          \label{fig:moments}
}
\end{figure*}

In Figure \ref{fig:moments}, we plot the theoretical bias for moments of
the density field in our simulations together with predictions from
Equation \ref{eq:moments} for $n=2,3,$ and $4$.  We see that even
though $\delta^4$ becomes non-linear at $R>2$ Mpc/$h$, the bias
predictions work surprisingly well down to $R\sim 0.3$ Mpc/h. In fact,
as long as the variance of the field is not-significantly larger than
unity, the predictions for $\delta^n$ bias are accurate, despite
$\left<\delta^n\right>$ being large, presumably due to isolated
high-density regions.

\subsection{Taylor expansion approximation to the $\tau$ field}
\label{sec:tayl-expans-appr}

We start by a simple test, that is demonstrating the hopelessness of using a 
Taylor expansion in modeling the flux fluctuations. Namely, we can expand
flux as
\begin{equation}
  F=\exp\left(-\bar{\tau}\left(1+\delta_\tau\right)\right) =
  \exp\left(-\bar{\tau}\right) \left[ 1- \bar{\tau}\delta_\tau +
    \frac{1}{2}(\bar{\tau}\delta_\tau)^2 - \ldots\right]
\end{equation}
and then directly test the applicability of this expansion in
simulations, by using a truncated Taylor expansion in place of an
exponential when creating the flux field in simulations, but then
proceeding to calculate biases as usual. Discrepancies between the 
truncated exponential expansion and the true field will show what is the
best any analytically method based on Taylor-series expansion could do.

\begin{figure*}
\centering
\begin{tabular}{cc}
   \includegraphics[width=0.5\linewidth]{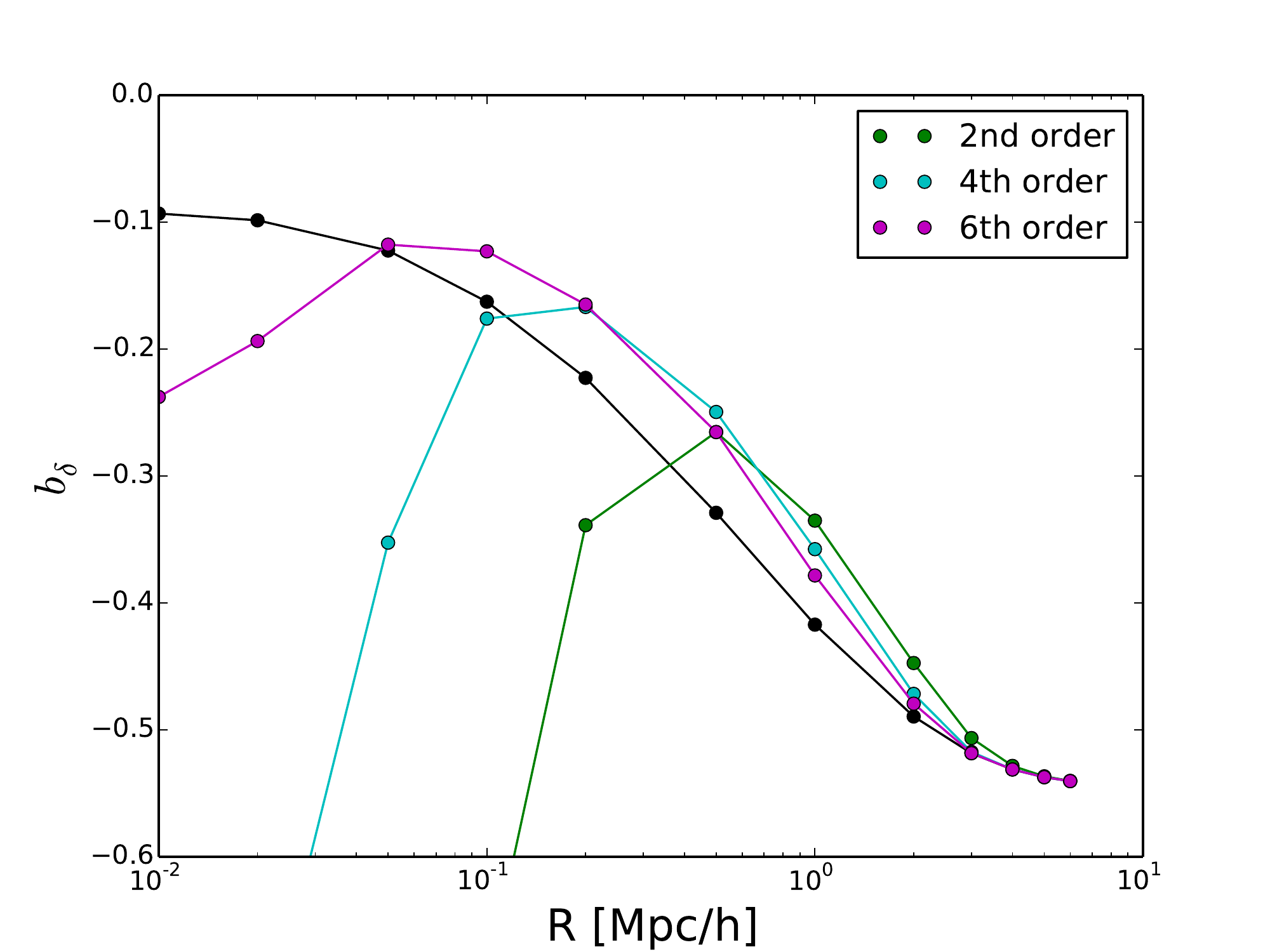} 

\end{tabular}
	\caption{Density bias of the flux fluctuations after calculating the flux field as a Taylor expansion in optical depth around its mean (see text). The black line represents 
	the full flux calculation (no Taylor expansion). The green
	line represents the Taylor expansion to 2nd order, the cyan represents 4th order, and magenta represents
	a 6th order expansion. All flux calculations are done for real-space FGPA with $A=0.34$ and $\alpha=1.6$ for $z=2.5$ using
	the peak-background split method described in the text.
          \label{fig:taylorseries}
}
\end{figure*}

Results for the simple real-space FGPA bias are shown in Figure \ref{fig:taylorseries}. We
see that the expansion works fine at large smoothings, but fails to
describe reality at small values of smoothing kernel size. This
demonstrates that any approximations that attempt to describe the
\lyaf\ bias in terms of bispestrum, trispectrum, etc of the $\tau$
field are bound to fail. What really drives the change of the bias is the
change in the flux PDF in the presence of the large-scale mode.

\subsection{Real-space FGPA}

\begin{figure*}
  \begin{tabular}{cc}
\includegraphics[width=0.5\textwidth]{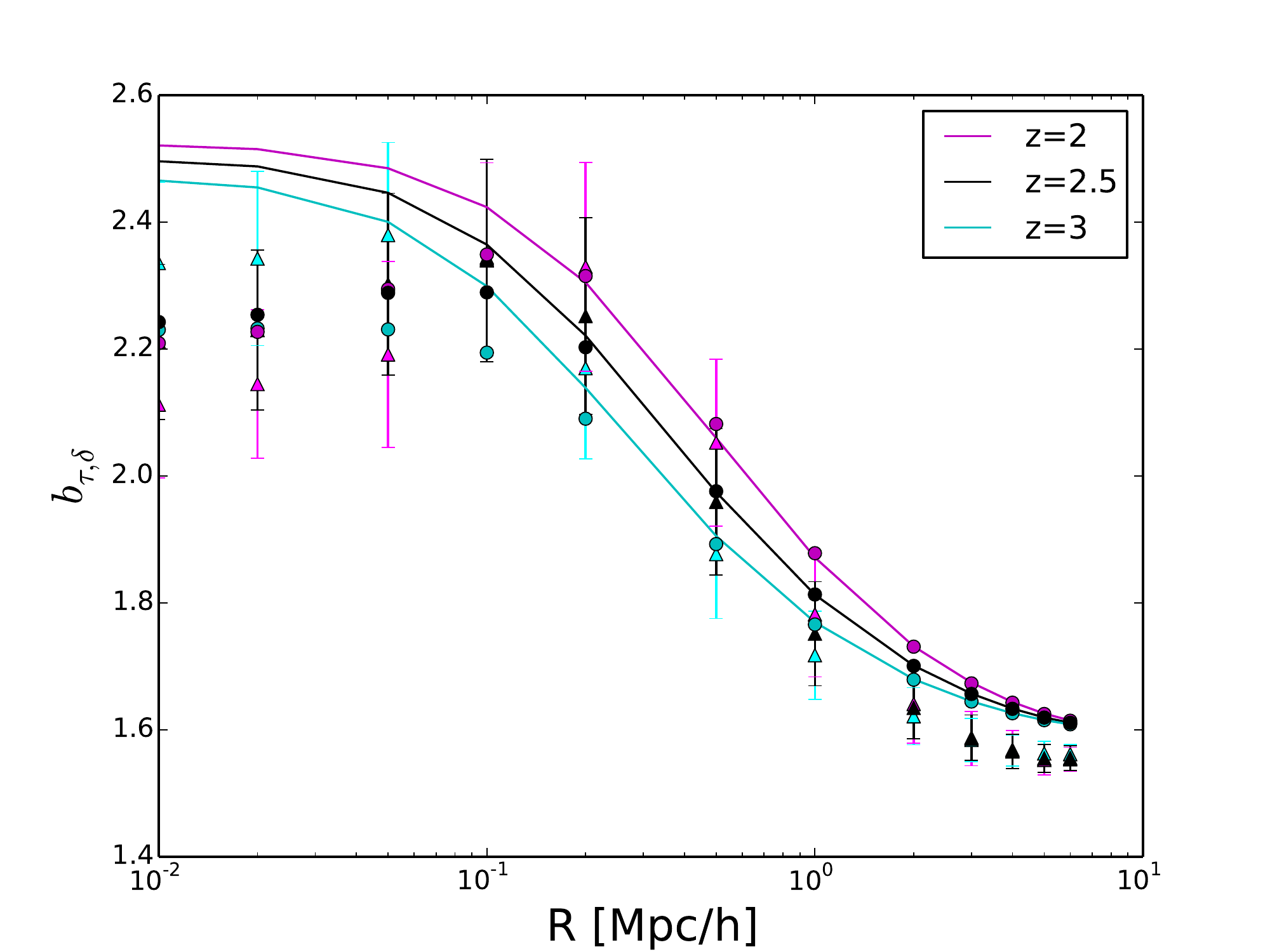}  &
\includegraphics[width=0.5\textwidth]{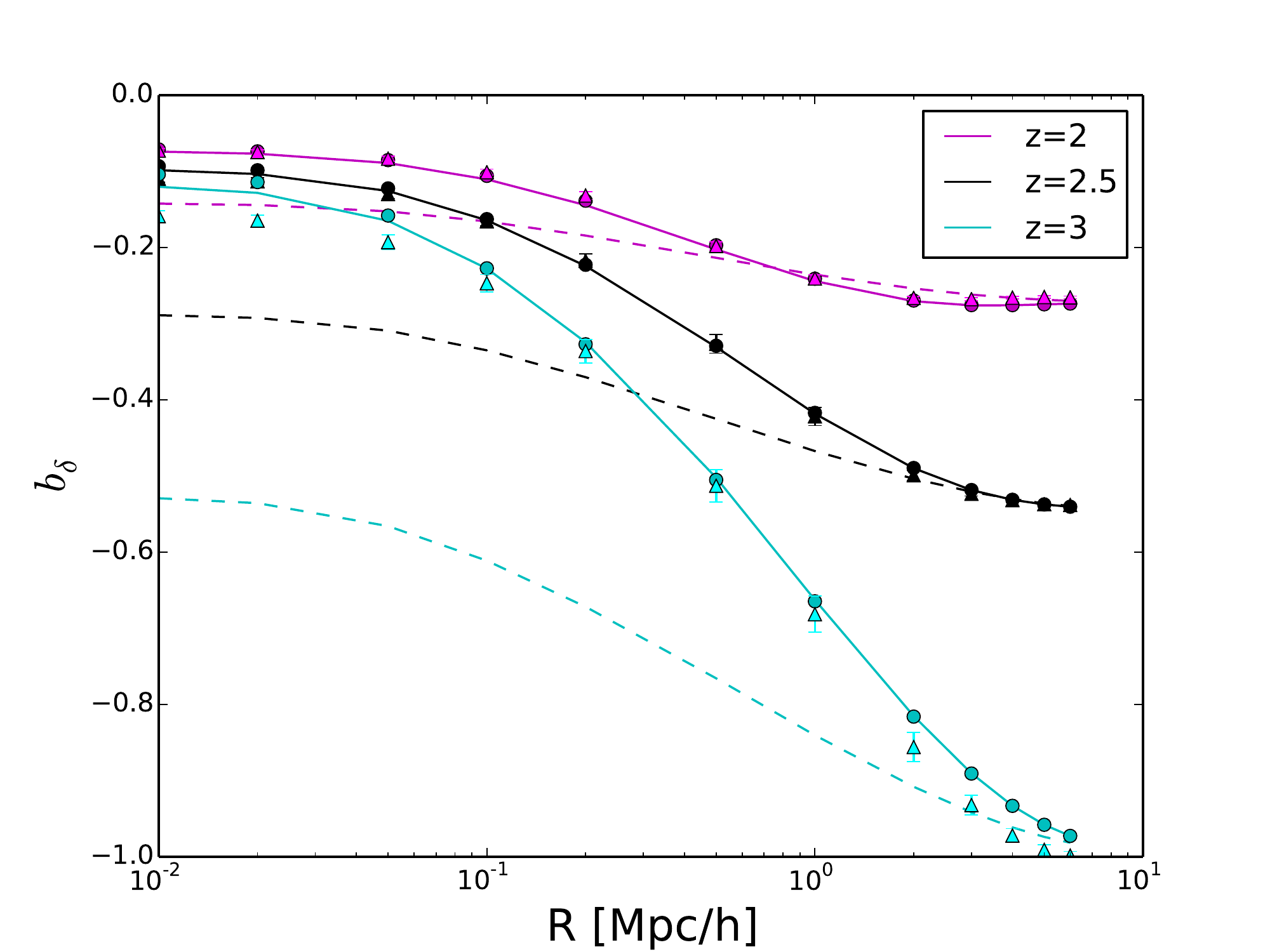} 
\end{tabular}	

\caption{Measured and predicted values of $b_{\tau,\delta}$ (left
 panel) and $b_{\delta}$ (right panel) in real space as a function of
  smoothing scale for $z=2$ (magenta, top), $z=2.5$ (black, middle) and $z=3$
  (cyan, bottom). Solid lines are analytical formula predictions, solid points
  are PBS results and triangles mode fitting. Dashed lines show
  analytic predictions with $\nu_2=0$.}
\label{fig:realspace}
\end{figure*}

In Figure \ref{fig:realspace} we plot the corresponding predictions
for $b_{\tau,\delta}$ and $b_\delta$ for real-space $\tau$. Analytical
predictions agree remarkably well with measurements for this deterministic
relation and are in fact much better than one would naively expect given
Figure \ref{fig:moments}. We find that the agreement for flux is
better than the agreement for the optical depth field. This is likely
because the $\exp(-\tau)$ transformation suppresses the high-density
regions where linear theory is supposed to break down.

The agreement between the theory and the measurement hinges on the
transformation of the flux PDF on the modes as given by the second order
linear calculation in Equation \eqref{eq:1}, which one would naively
expect to be swamped by higher-order non-linear effects on small
scales. This good agreement is not a coincidence.  If we set $\nu_2=0$ (dashed line
in Figure \ref{fig:realspace}) we get completely wrong results, so the
magic number $\nu_2=34/21$ stemming from second order standard
perturbation theory is indeed crucial to the good fit we obtain
in this case. As demonstrated in section \ref{sec:tayl-expans-appr},
this success does not stem from the second order perturbation theory
correctly capturing the bispectrum of $\tau$ fluctuations, but rather, it is
because Equation \eqref{eq:1} correctly captures how the flux PDF is
transformed in the presence of the large-scale mode.

We also see that the analytical prediction works increasingly well
with decreasing redshift, in correspondence with the decreasing value
of $A$. With no smoothing ($R=0$), the analytical prediction is within
$4\%$ of the peak-background split results for $z=2$, where $A=0.17$,
within $6\%$ for $z=2.5$, where $A=0.34$, and within $18\%$ for $z=3$,
where $A=0.62$. The analytical prediction therefore works best for low
values of the optical depth, which decreases with decreasing redshift,
despite the increase in the matter power spectrum. On the other hand,
if $A$ is held artificially constant between redshifts at $A=0.34$,
the corresponding values are $16\%$ for $z=2$ and $7\%$ for $z=3$, as
expected, for linear theory to hold better at higher redshift. If the
density field is smoothed with a smoothing factor of $R=0.2$ Mpc/h,
thus decreasing the overall value of the optical depth, the analytical
prediction is still within $4\%$ of the peak-background split results
for $z=2$, where $A=0.17$, but now $0.6\%$ for $z=2.5$, where
$A=0.34$, and within $0.8\%$ for $z=3$, where $A=0.62$.

The slightly worse agreement between the numerical and analytical methods at higher redshift is therefore driven by the increasing value of the normalization, $A$. The IGM is less ionized at these redshifts, resulting in a higher optical depth. The higher $A$ therefore may pick out the few extreme regions of the simulation, skewing the bias measurement towards these regions. Since the sample size of these regions inside the volume being probed by our simulations is small, the error on this measurement will be higher at larger redshift, limited by this sample variance. 

\subsection{Redshift-space FGPA}

We introduce the redshift-space distortions in three different
ways of increasing realism to test how well do theory predictions
fare. 

\subsubsection{Linear Kaiser RSD}
 First, we impose the redshift-space distortions by simply adding
them to the real-space $\tau$ FGPA field as expected in the linear
theory:
\begin{equation}
\tau_s = \tau_r (1+f \mu^2 \delta_l)
\end{equation}
This is justified by the fact that the $\tau$ field is conserved on
real-redshift-space transformation and hence $b_{\tau,\eta}=1$.  We
refer to fields created in such a way as ``linear Kaiser RSD''
(lKRSD).\footnote{ Note that one cannot introduce redshift space
  distortions using $\delta_{\tau_s} = \delta_{\tau_r} + f \mu^2
  \delta_l$, since this drops the cross terms, which are not
  negligible, and produce widely disparate results for smaller
  smoothing scales than the formula above.}

\subsubsection{Linear velocity RSD}

Next we produce an actual redshift-space $\tau$ by shifting the $\tau$
field around using appropriate velocities, following
\begin{equation}
\tau_s(s) = \int \tau(r) \delta_D(s-r-H^{-1}v) dr,
\label{eq:rsdmove}
\end{equation}
but actually using linear velocities $v_k = i f a H(a) \delta(k) \mu_k
/k$ for this. We refer to this method of generating RSD as ``linear velocity RSD" (lvRSD).

\subsubsection{Non-linear velocity RSD}
Finally, we use equation \ref{eq:rsdmove}, but use actual non-linear
baryon velocities. We refer to this method of generating RSD as
``non-linear velocity RSD'' (nlvRSD).

\subsubsection{Thermal Broadening}
In addition to the velocity offset in a simple RSD picture, the
absorbing neutral hydrogen gas also has some thermal motion with
respect to the background due having a gas temperature $T$. This
additionally introduces a Gaussian line broadening profile:

\begin{equation}
\sigma_\alpha=\sigma_0 \frac{c}{b \sqrt{\pi}} e^{-(\Delta v)^2/b^2},
\end{equation}
where $\sigma_\alpha$ is the new cross section written in terms of the
original \lya\ absorption cross section at rest, $\sigma_0$, broadened
by a Maxwellian velocity distribution of the gas with a Doppler
parameter $b=\sqrt{2 k T/m}$, where $m$ is the mass of the atom, and
$\Delta v$ is the velocity difference with respect to the center of
the absorption line.

In effect, RSD and thermal broadening together move the optical depth
by a corresponding peculiar velocity and spread out the value at that
point with a Gaussian profile to neighboring pixels with a width based
on the temperature of the gas at the original point. We therefore
expect thermal broadening to be degenerate with the $b_\eta$ value of
the pure RSD, as these are both line-of-sight effects.

We use the tight temperature-density relation to introduce this
thermal broadening, with $b^2 = (12.8 \mathrm{km} \mathrm{s}^{-1})^2
\left(\frac{T_0}{10^4 K}\right)(\delta+1)^{\gamma-1}$, and
$\gamma-1=(2-\alpha)/0.7$ and $T_0=10^4K$.

\subsubsection{Results}

For the peak background split model calculations, we rescale and
smooth the peculiar velocities just as the RSD density in the previous
section (with an additional factor of $f$). We use this rescaled
velocity field for the peak-background split calculation for the
simple redshift space fields.

\begin{figure*}
  \begin{tabular}{cc}
    \includegraphics[width=0.5\textwidth]{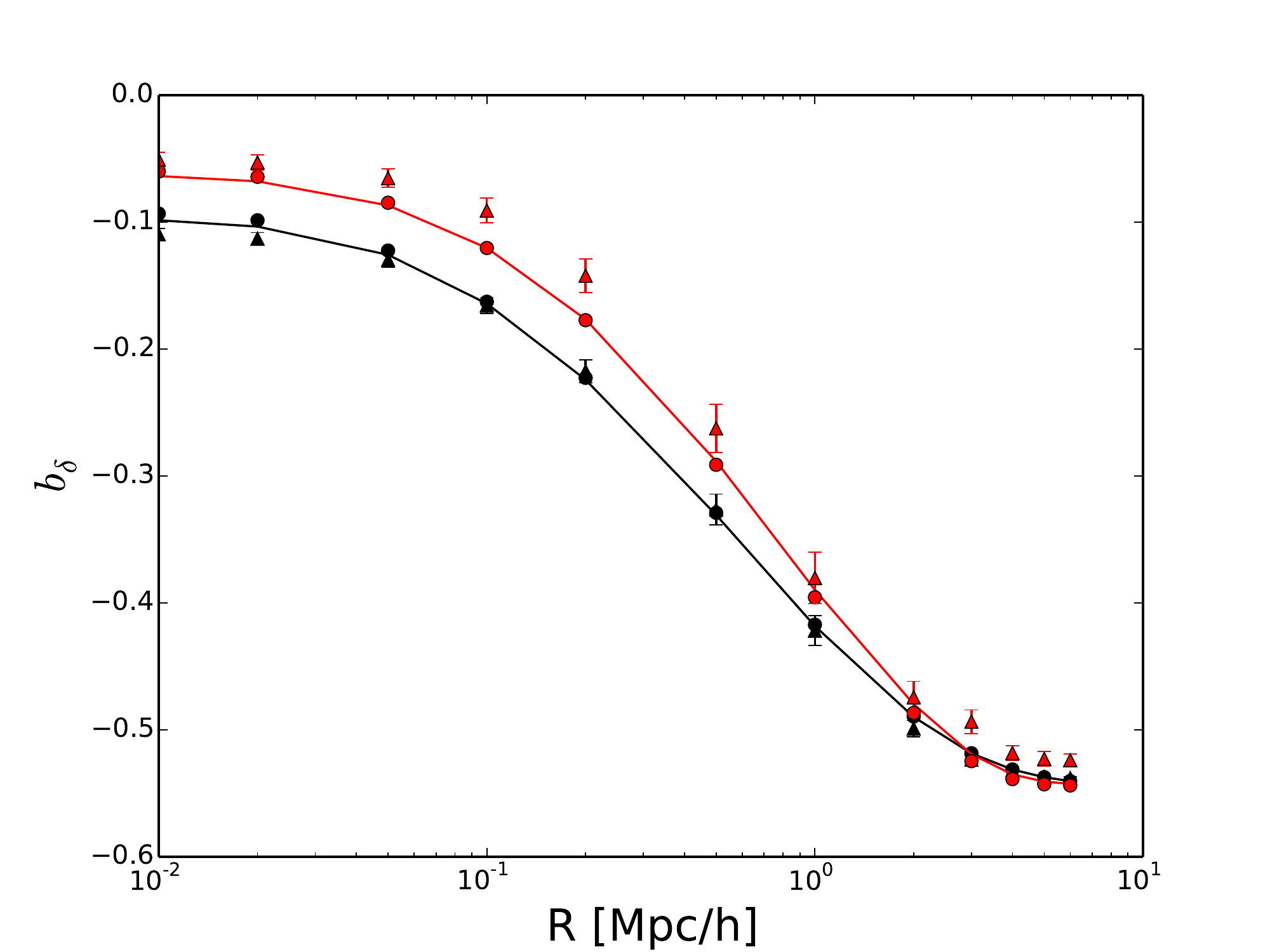} &
    \includegraphics[width=0.5\textwidth]{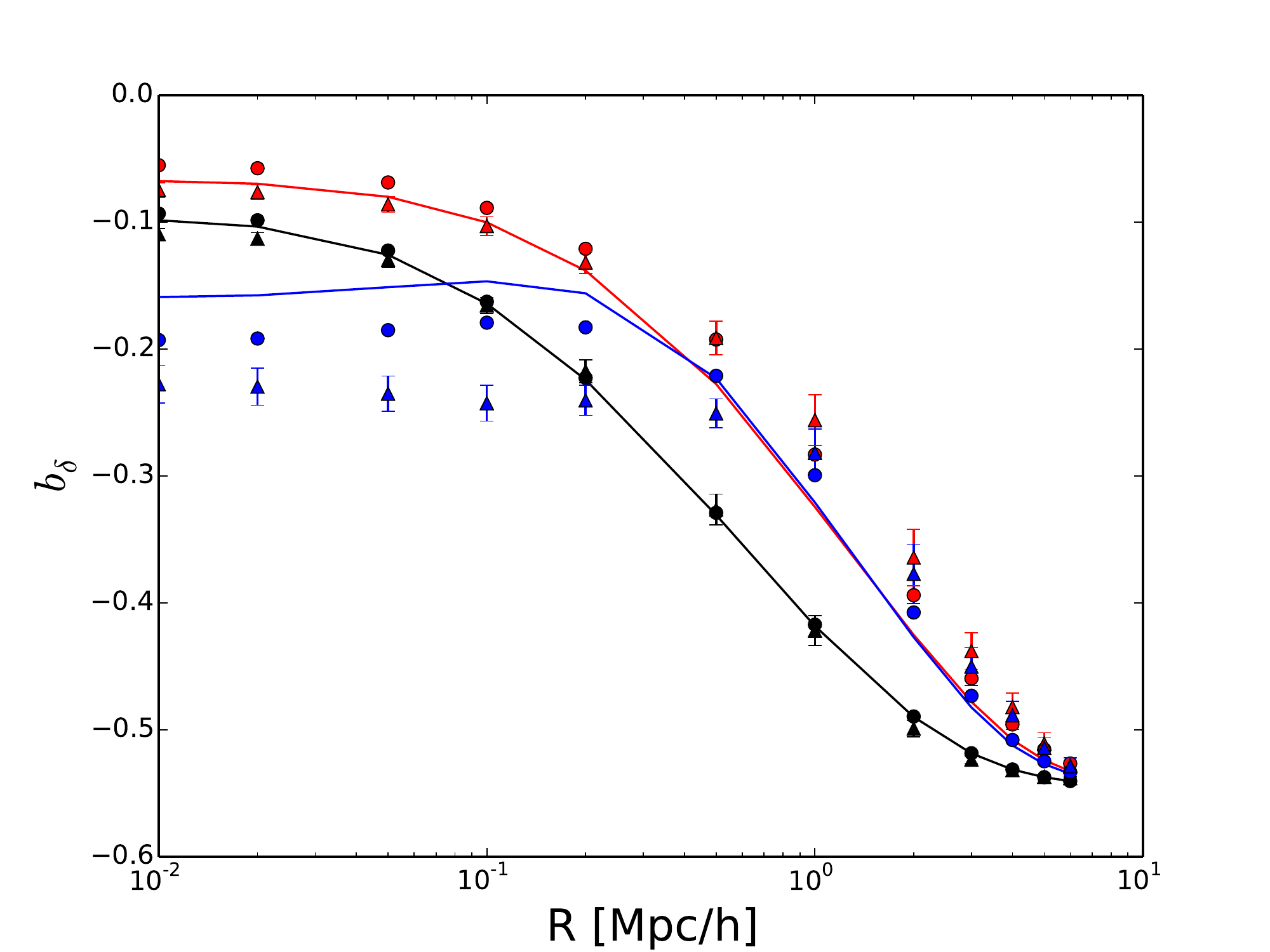} \\
    \includegraphics[width=0.5\textwidth]{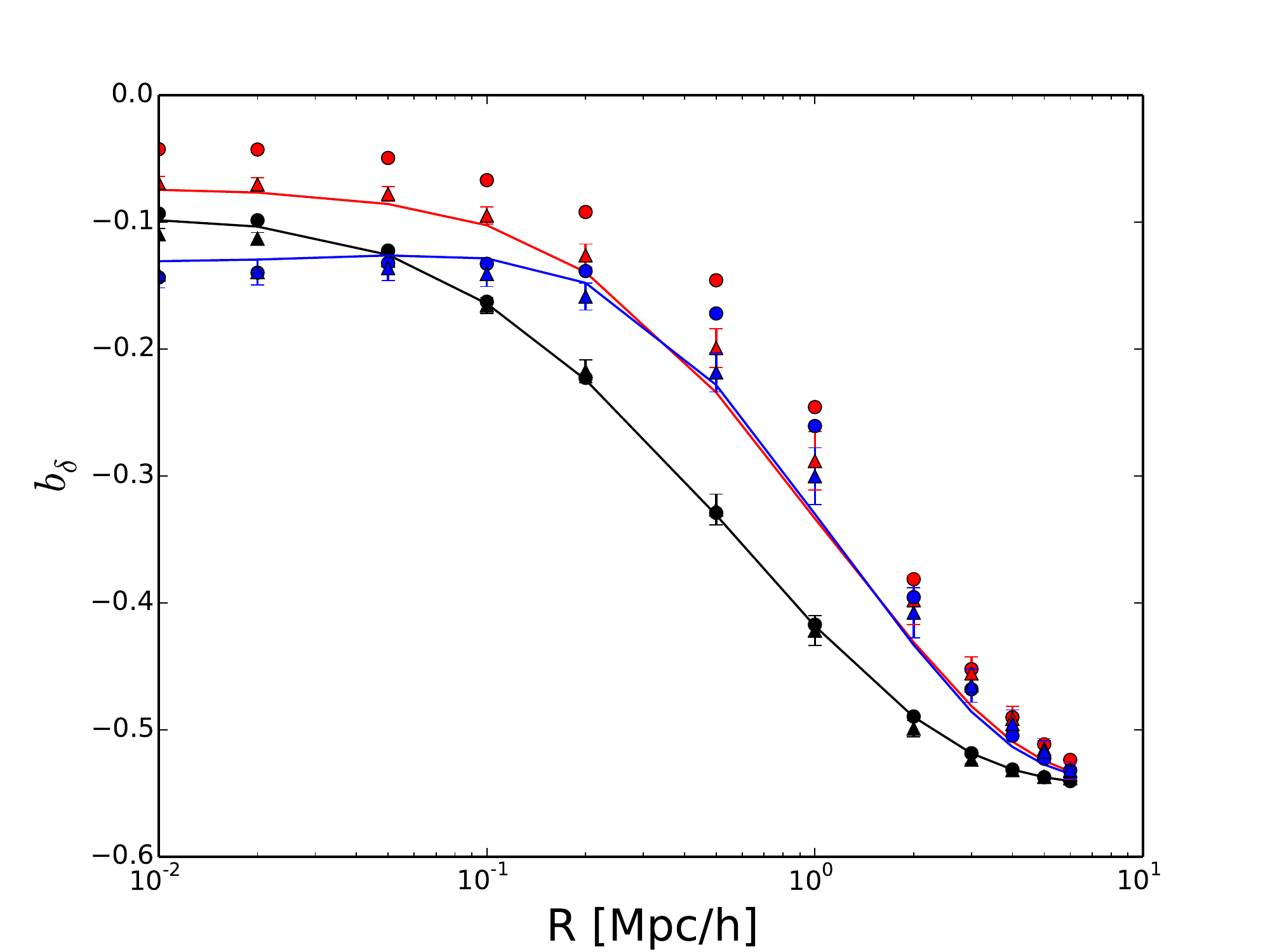} &
    \includegraphics[width=0.5\textwidth]{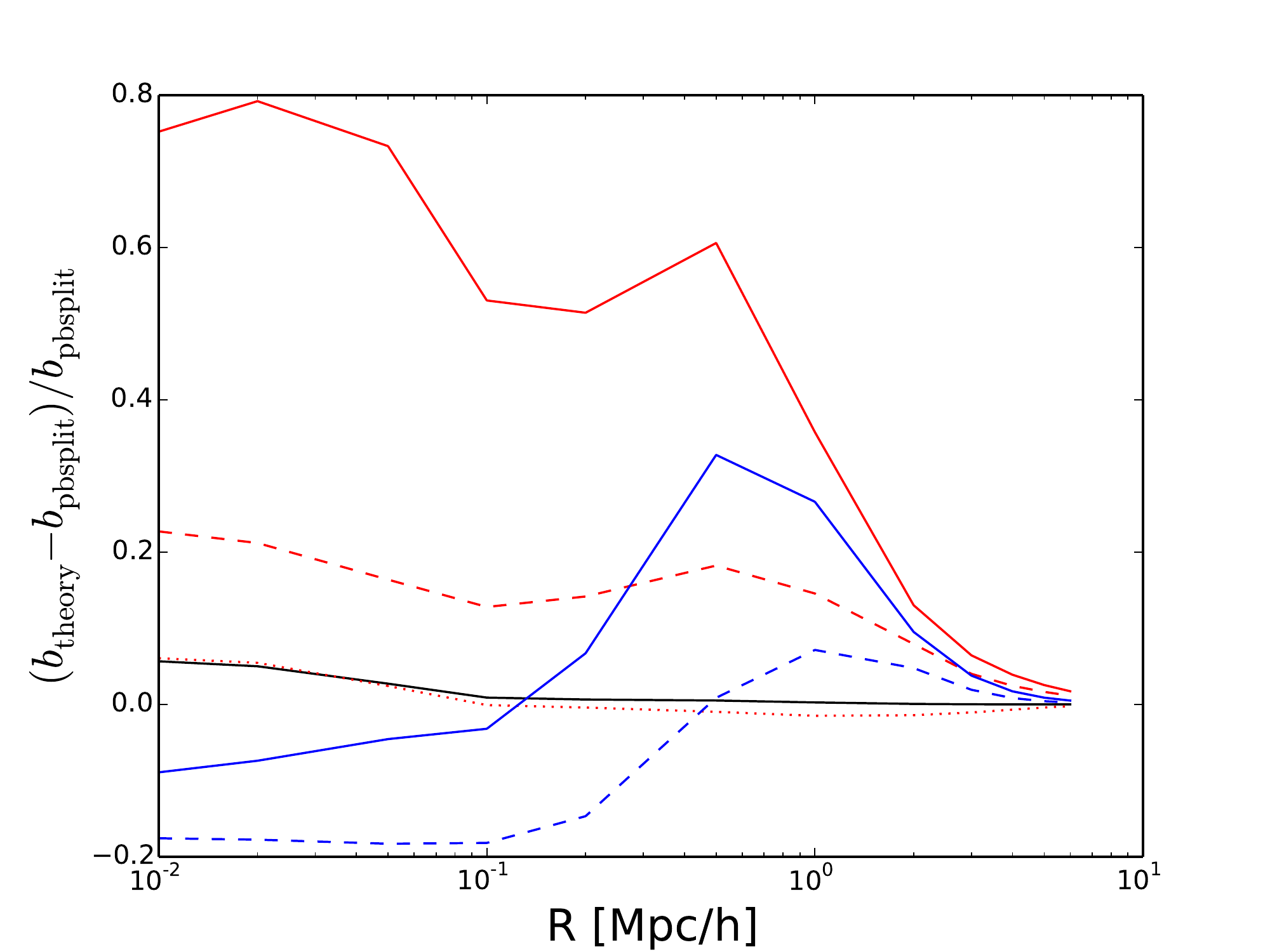} \\   
  \end{tabular}
  \caption{Flux bias parameter for $z=2.5$ and $A=0.34$ and
    $\alpha=1.6$, while varying the smoothing parameter $R$. Black
    represents real-space, red redshift space, and blue introducing
    redshift space and thermal smoothing, as described in the
    text. Lines, points and errorbars correspond to theory, PBS method
    and likelihood methods of bias determination. The top left panel
    corresponds to lKRSD, top right to lvRSD and bottom left to nlvRSD
    (see text for description of acronyms). The bottom right panel
    shows the difference between the analytical predictions and the
    PBS method. The dotted lines represent lKRSD, dashed lvRSD and
    solid nlvRSD. }

\label{fig:rsdbias}
\end{figure*}

Our results for the  flux density bias are displayed 
in Figure \ref{fig:rsdbias} in red.
We see that as we progress towards more realistic redshift-space
distortions, the theory predictions for the density bias remain in rough
qualitative agreement with the measurements, but that the formula is
accurate only to within an order of magnitude. Looking first at the
PBS results, we note that while purely Kaiser distortions (lkRSD) seem to
keep the predictivity of our formulae, even taking into account the
non-linearities induced by actually moving the $\tau$ field around
(instead of imposing $\mu^2$ distortions) already introduces errors at
the level of 20\%, which then increase even further when we go to correctly non-linear
velocities. Adding thermal broadening correctly captures the decrease
in bias at small smoothing, but underpredicts the overall effect. When
thermal broadening is applied on nlvRSD, the formula is accurate to
around 10\% for $R=(10^{-2}-10^{-1})$Mpc/$h$, but this seems to be
mostly due to fortuitous cancellation of errors (since it works worse
in the case of lvRSD). We also note a surprising discrepancy between the
likelihood method for measuring bias and the PBS in the case of nlvRSD
and no thermal broadening. This can be directly tracked to the fact
that there are missing velocity non-linearities in the 40 Mpc/h box that
cause the offset in PBS measurements with respect to the theory
predictions -- if we replace the velocity field in the offset simulations
with that of the original simulation, the difference between
likelihood and PBS biases disappear.

\begin{figure*}
\begin{tabular}{cc}
    \includegraphics[width=0.5\textwidth]{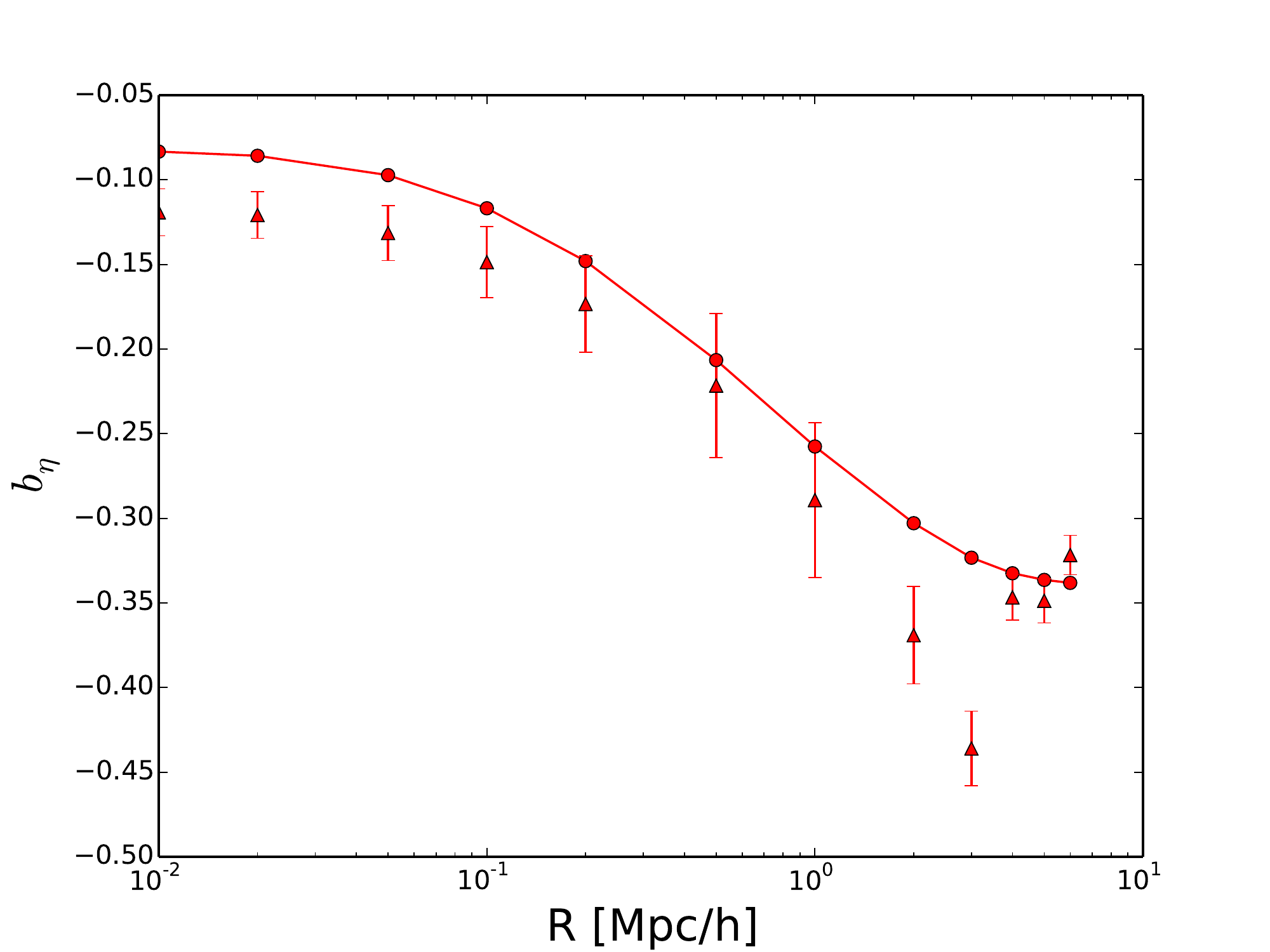} &
    \includegraphics[width=0.5\textwidth]{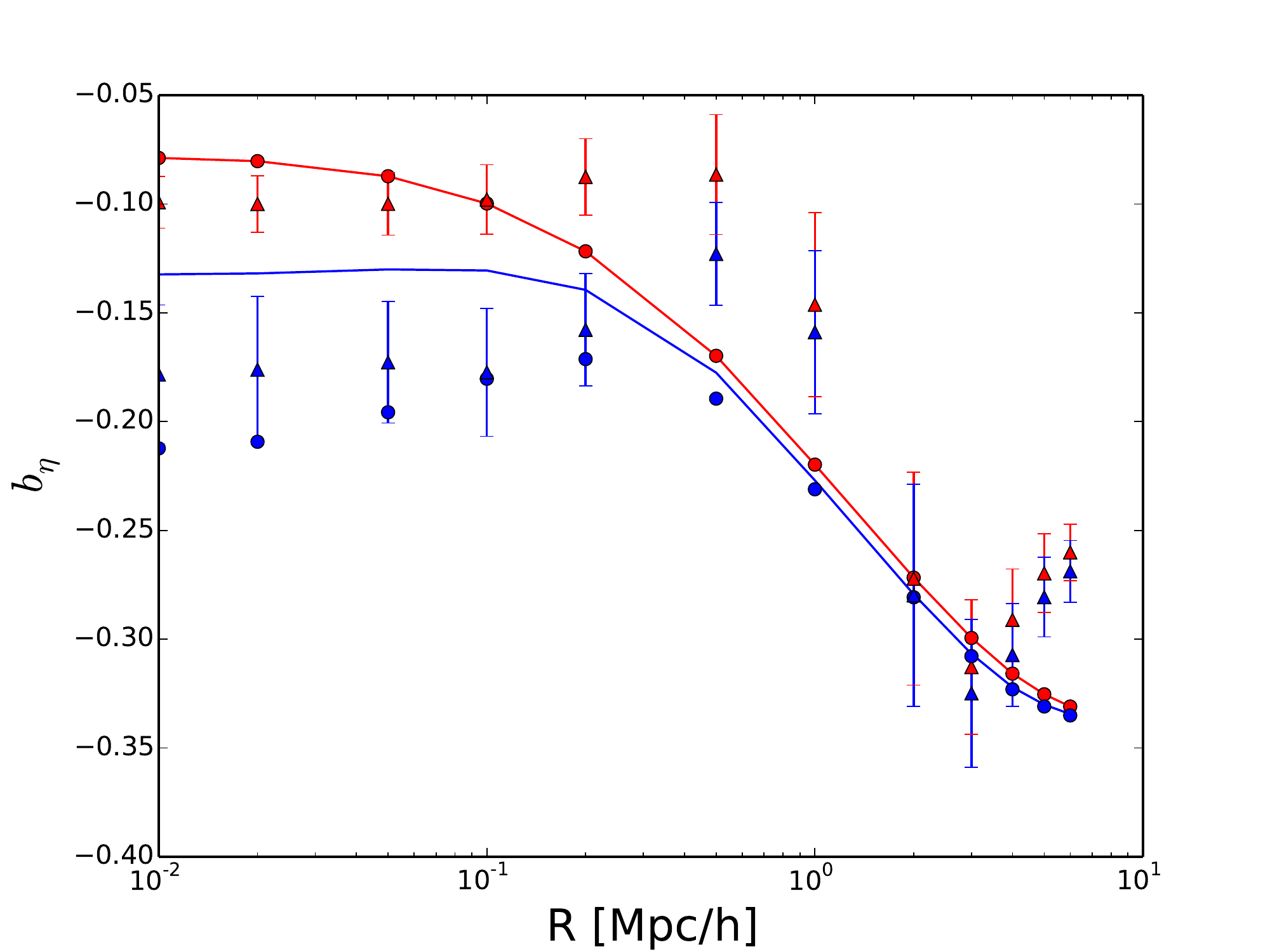} \\
    \includegraphics[width=0.5\textwidth]{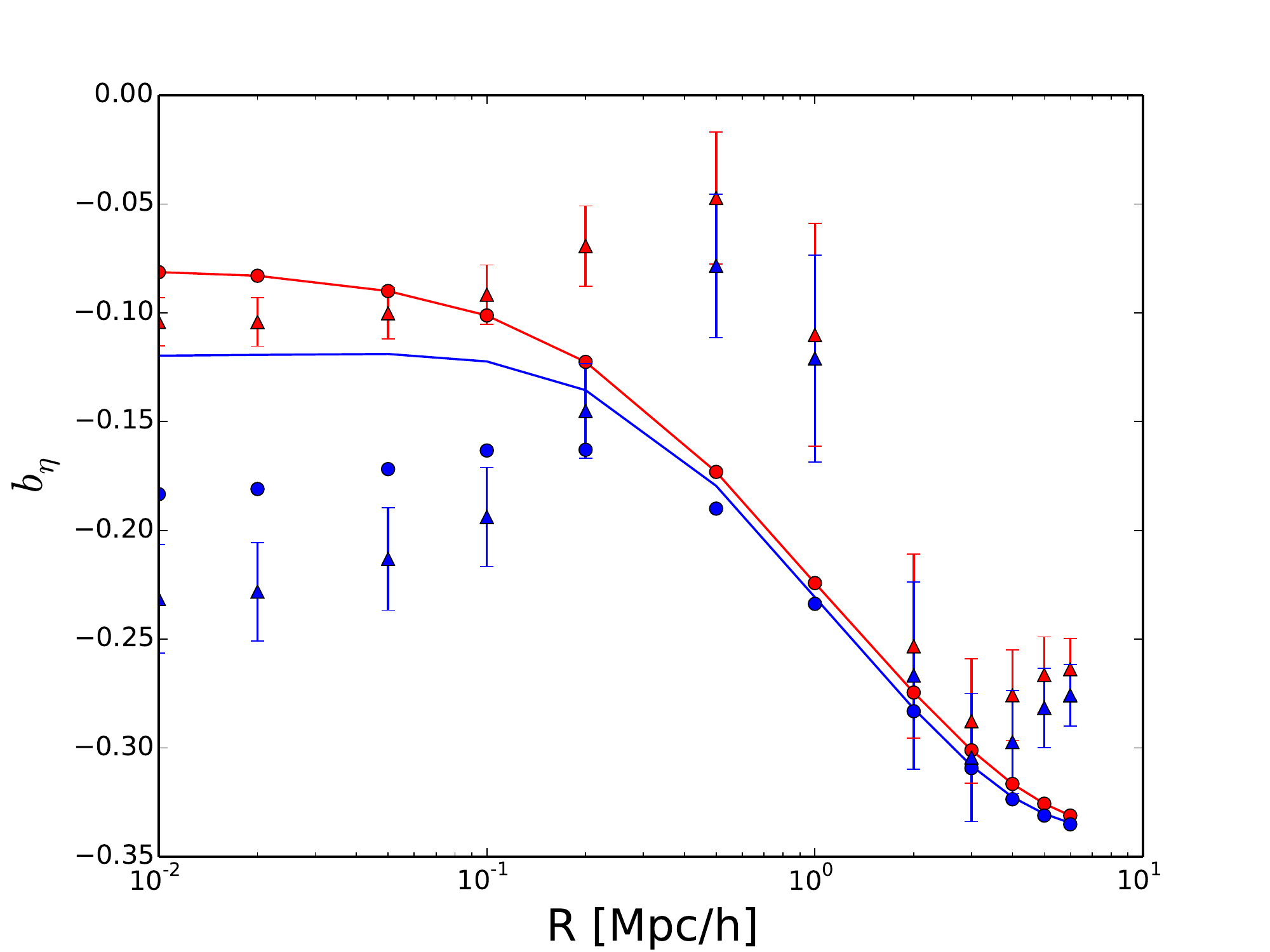} &
    \includegraphics[width=0.5\textwidth]{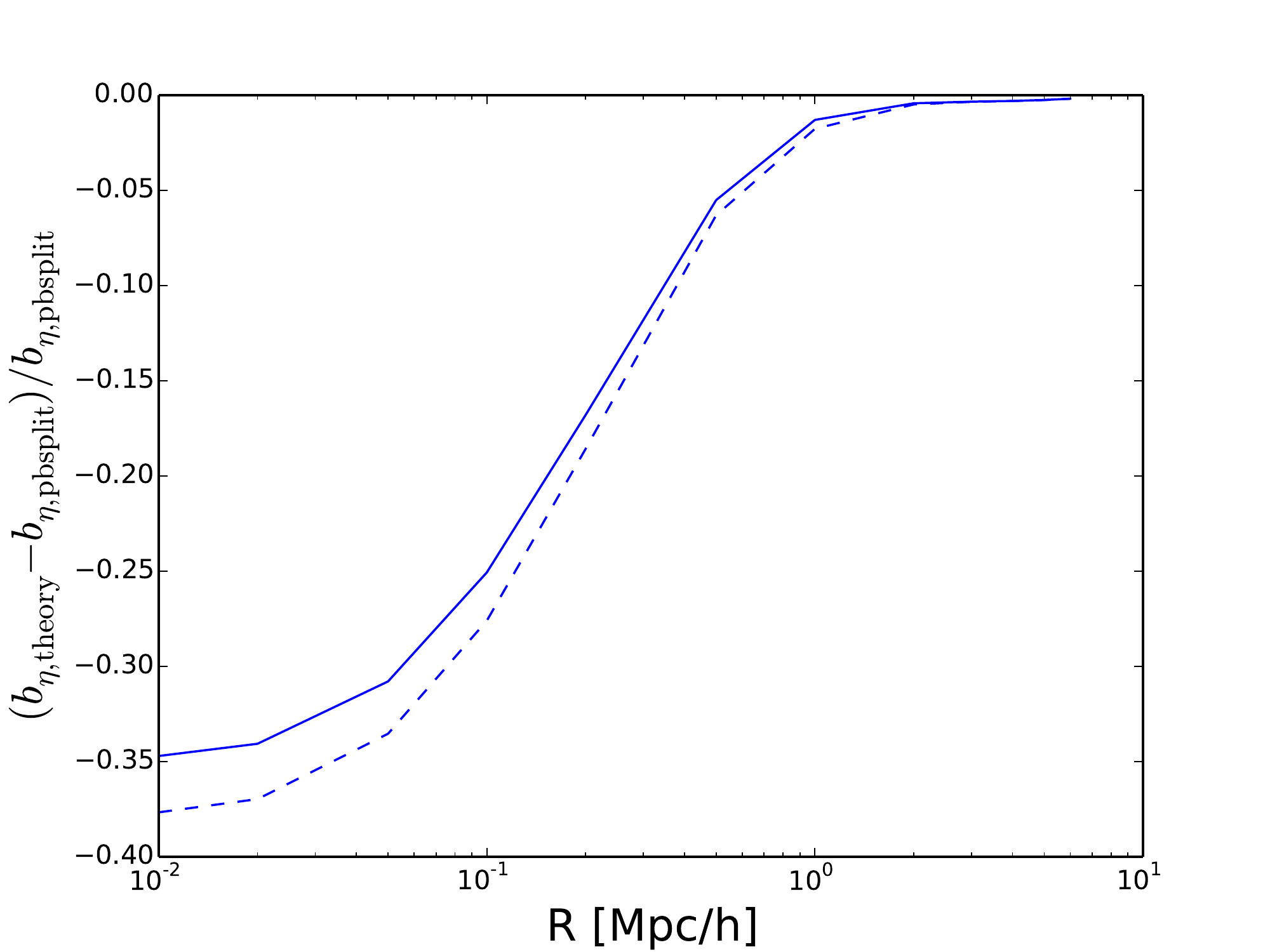} \\
  \end{tabular}
  \caption{Same as Figure \ref{fig:rsdbias} but for $\bieta$.}
\label{fig:rsdbeta}
\end{figure*}

Figure \ref{fig:rsdbeta} shows the same for the velocity gradient
bias. In this case, in case of no thermal broadening, the theory
predictions are exact and show perfect agreement with the PBS method. Note
however that the introduction of thermal broadening has a devastating effect
on the accuracy of this prediction, making it underestimate the bias
by up to 30\%.  However, we note that this difference could actually
turn out to be a sensitive probe of the thermal broadening and thus
the temperature of the IGM, but we defer this study for future work.
Again we note curious discrepancies between the power spectrum bias
predictions and the PBS, which are occasionally larger than what one
would expect based on errors alone. Our tests indicate that this is
unlikely due to underestimation of errors, but can be due to missing
$\eta$ bias parameters and simply small box effects. 

\section{Dependence of $A$ and $\alpha$}

\begin{figure*}
  \begin{tabular}{cc}
    \includegraphics[width=0.5\textwidth]{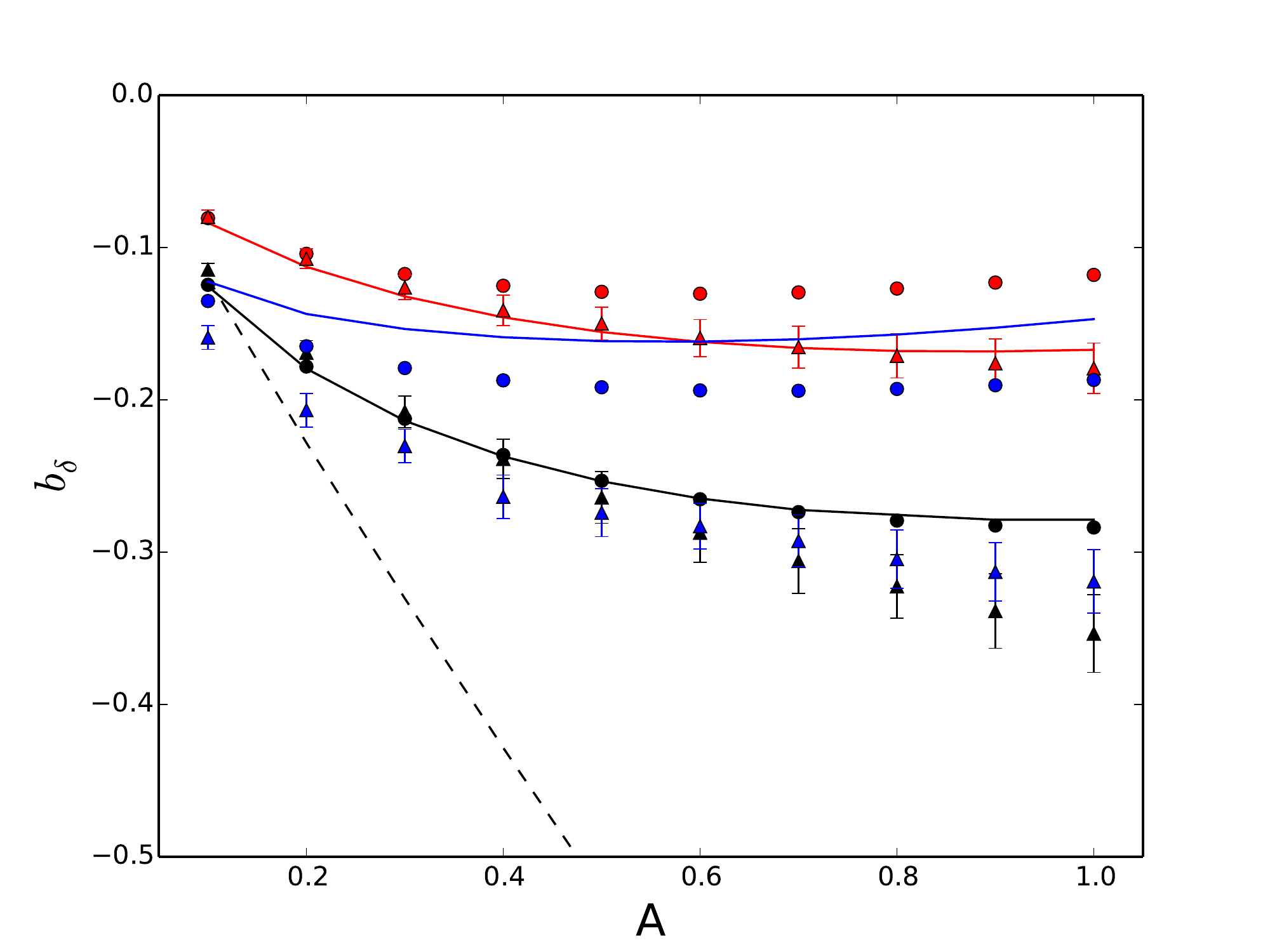} &
    \includegraphics[width=0.5\textwidth]{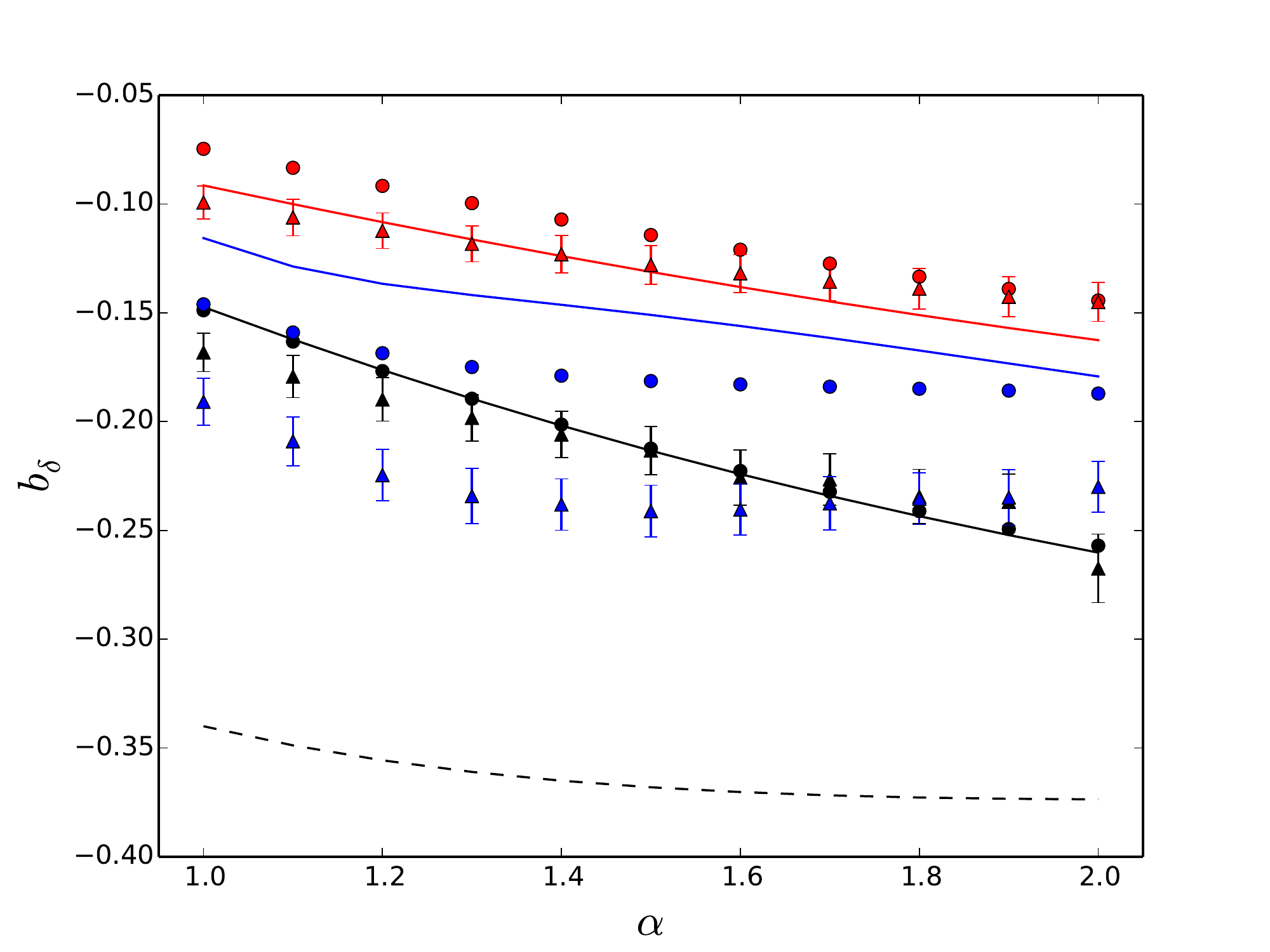} \\
  \end{tabular}
  \caption{ Flux bias parameter for $z=2.5$ and $R=0.2$ Mpc/h, while
    varying the parameter $A$ at constant $\alpha=1.6$ (left panel) and
    varying $\alpha$ at constant $A=0.34$ (right panel) for lvRSD.  Black
    represents real-space, red redshift space, and blue introducing
    redshift space and thermal smoothing. As before, solid lines are analytical formula predictions, solid points are PBS results and triangles mode fitting. Dashed lines show analytic predictions with $\nu_2 = 0$.
     }
\label{fig:biasvsA}
\end{figure*}

Figure \ref{fig:biasvsA} shows the dependence on the values of $A$ and
$\alpha$ for a smoothing scale of $R=0.2$ Mpc/h for $z=2.5$. The
agreement of the second-order analytical prediction of
\cite{2012JCAP...03..004S} with the peak-background split method in
real-space is to within $10\%$ at this central redshift for a wide
range of values for $A$ and $\alpha$. Again, a first-order
perturbation theory with $\nu_2=0$ fails miserably.

However, after the redshift-space distortions are introduced, both
methods seem to fail at about the same level, regardless of values of
$A$ and $\alpha$ and only weakly capture the trends. 

Again, we note an interesting discrepancy between PBS bias
determination and the power spectrum based one, increasing towards
large values of $A$. These correspond to more aggressively picking
particular regions of the space where finite volume effects play a more
important role.

\section{Hydrodynamical fields}

We now turn to the full hydrodynamic part of the simulation, taking the flux field
generated directly from the simulation. We rescale the flux field to match the mean 
values of \cite{2013MNRAS.430.2067B}. The analytical and likelihood bias methods are calculated directly
from these fields, where the values of $A$ and $\alpha$ are taken from a smallest chi-squared 
fit of the real-space $\tau$ field with respect to the underlying density field.

However, for the peak-background split method, the situation is a bit more complex.
We note that while the density field in the perturbed simulations evolve as expected with respect to the 
fiducial simulation, the ionizing background is evolved with the redshift of that simulation, not corresponding
to the time of the fiducial simulation. We therefore have to rescale the fitted value of $A$ to fit that of the fiducial
simulation at the same corresponding global time. This is further complicated by the fact that we do not
have outputs of these perturbed simulations that would correspond to the exact same global time as those of
the fiducial simulation (we do not have outputs at the values of $z'$ as stated in Section 3.1). In order to circumvent this, 
we first interpolate (with a polynomial fit) the rescaling of $A$ in the fiducial simulation with respect to redshift, we then
calculate the redshift of the fiducial simulation that would represent the same global time as the output of the perturbed simulation 
(for example, $z'=2.5$ corresponds to a $z=2.53$ in the fiducial), and rescale the optical depth at that output according to the 
rescaling function of $A$ (in the example, this would correspond to rescale the perturbed simulation to match A(z=2.53)).
After rescaling the optical depths to the appropriate values of $A$ at the output redshifts, we calculate the mean values of flux, and then again
use an interpolating function for this mean flux as a function of redshift, to calculate the mean flux values at the appropriate values of $z'$ 
according to Equation A4. This method introduces an error due to the interpolating functions, but could be rectified in the future by generating outputs
of the perturbed simulations at the appropriate values of $z'$. 

\begin{figure*}
  \begin{tabular}{cc}

	\includegraphics[width=0.5\textwidth]{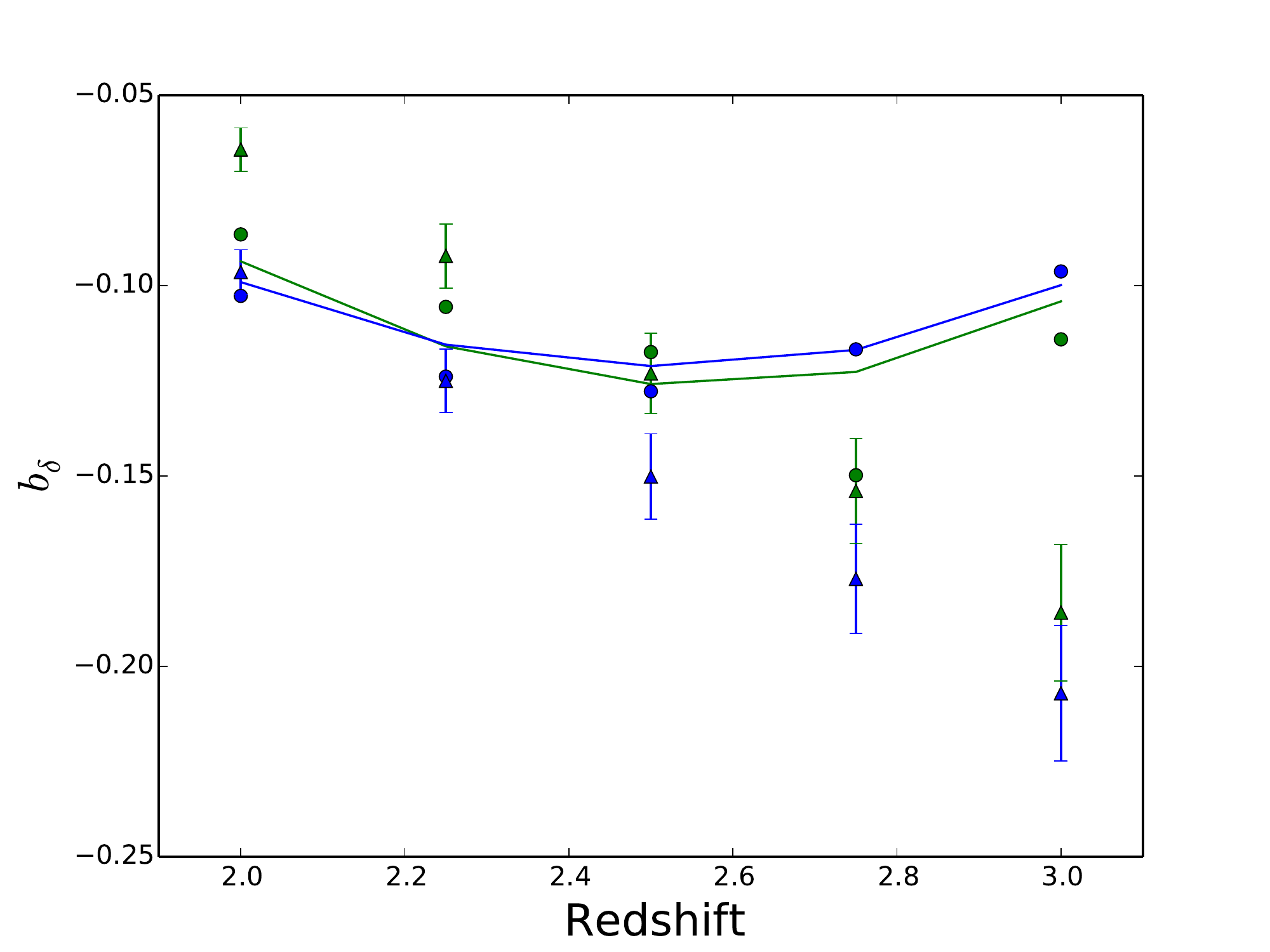} &
	\includegraphics[width=0.5\textwidth]{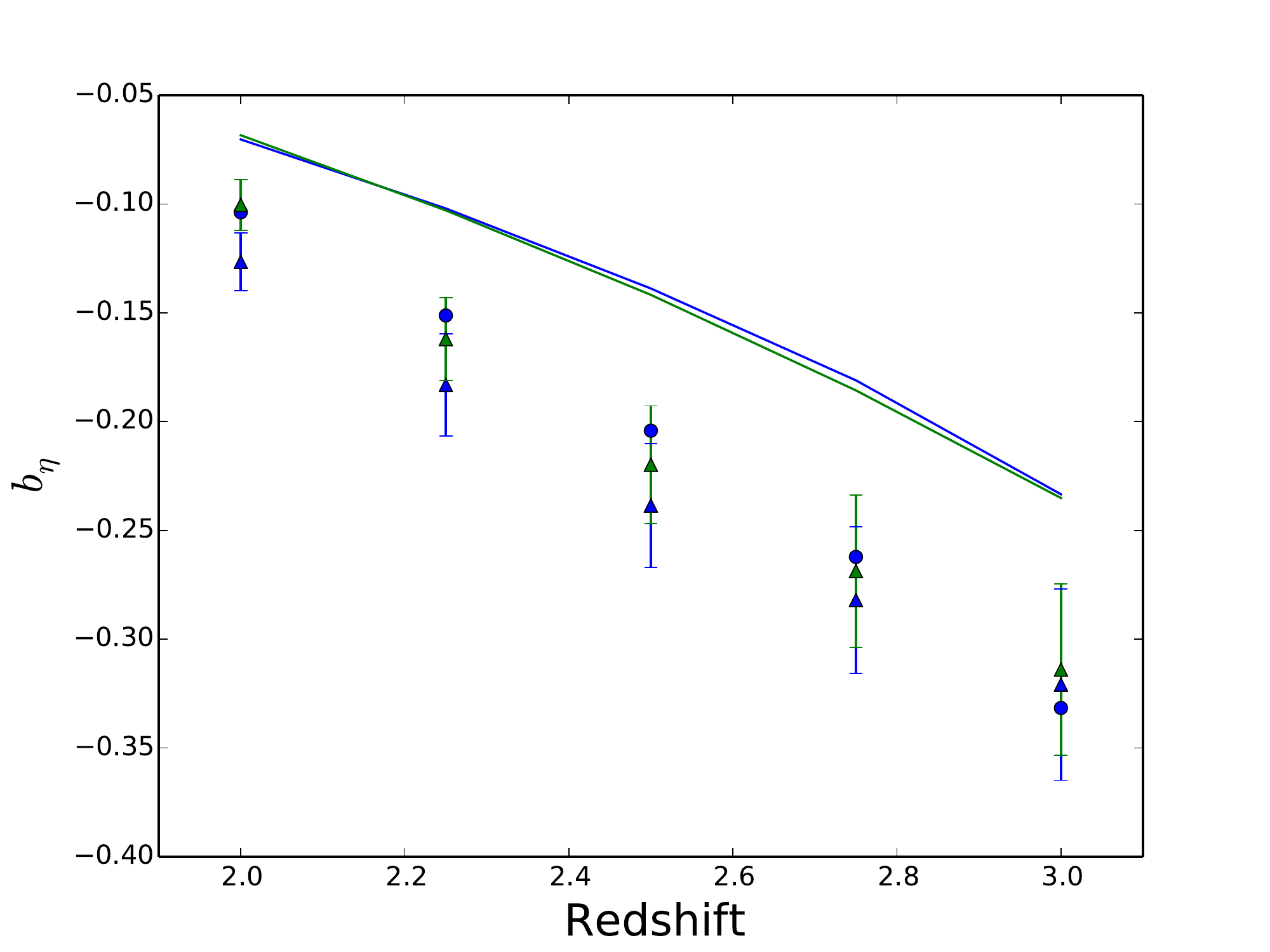} \\

       \end{tabular}

       \caption{Flux density bias (left) and velocity gradient bias
         (right) parameters for the hydrodynamic part of the
         simulation plotted in green with redshift, where the flux is
         scaled to match the means as measured by \cite{2013MNRAS.430.2067B}. As a comparison, the result of the FGPA + RSD + thermal
         smoothing method of the previous sections is plotted in blue
         for the corresponding $A$ and $\alpha$ as fitted in the
         simulation, and appropriately scaled to the mean with a
         smoothing of $R=0.04$ Mpc/h. The corresponding
         symbols have the usual meaning, where triangles are the
         likelihood method calculation, circles the peak-background
         split result, while the solid lines are the analytical
         predictions. }
\label{fig:hydrobias}
\end{figure*}

We compare this to the result of the FGPA + RSD (nlvRSD)+ thermal broadening method of the previous sections, with a smoothing 
value of $R=0.04$ Mpc/h, which corresponds to the value where the analytical predictions of the density bias in real-space flux fields were the closest
to those of the real-space hydrodynamic flux predictions. The comparison is plotted in Figure \ref{fig:hydrobias}.

It is hard to read too much into this figure, since the uncertainty in the
flux determinations is significant. We chose a value of smoothing so
that predictions based on the flux PDF are in closest agreement 
and these seem to be good to $10\%$.
The numerical value of the smoothing is in the right ball-park but it is
not the same as the filtering scale of
\cite{1997MNRAS.292...27H}.  We find
decent agreement between FGPA approximated fields and hydrodynamical
fields, especially at higher redshift.  

We find a curious upturn in $\bdelta$ with redshift, both in the PBS
determination of bias and the analytical predictions. This is likely the
result of errors in our bias determination and a failure of analytical
predictions at high values of $A$, since linear theory predicts a power-law 
change of bias with redshift \cite{2012JCAP...03..004S}. Observations of the bias trend
are consistent with our results for $2<z<2.5$ \cite{2011JCAP...09..001S}.

The above values calculated with the FGPA method also do not include scatter around the $\tau - \delta$ relation.
We investigate the effect of scatter by incorporating a scatter term into the relation, where $\tau=\tau_{\mathrm{FGPA}}(1+\epsilon)$, following 
\cite{2012JCAP...03..004S}, where $\epsilon$ is
taken from a normal distribution with a standard deviation of $\sigma_\epsilon=0.15$ as
measured from the hydrodynamic part of our simulations. This amount of scatter results in the FGPA analytical predictions
decreasing in absolute value by less than $1\%$ for both $b_\delta$, and
$b_\eta$ at the central redshift of $z=2.5$,  and is negligible for the numerical bias results.
The addition of scatter therefore is not an important effect in the comparison above.

For the velocity bias results, we are not able to generate the
peak-background split method result, but we present the likelihood
method and the analytical prediction, again comparing to the FGPA +
nlvRSD + thermal broadening methods in Figure
\ref{fig:hydrobias}. The hydrodynamic and FPGA fields are in good agreement with each other, but
while the analytical formula for both nicely predicts the relative change in $\bieta$ 
as a function of redshift, it fails to predict its absolute value. For both fields the absolute value of $\bieta$ is within
$27\%$ of the mode-by-mode fit for $z=2.0$, growing to a slightly worse agreement at $z=3.0$, where the absolute value of the analytical prediction is
within $37\%$ of the simulation mode-by-mode fit.

\section{Discussion \& Conclusions}

We have applied the FGPA framework in order to test the analytical
form for the flux density bias and the velocity gradient bias against
the numerical methods of the peak background split numerical
derivative and the mode-by-mode likelihood fit. We have found that the
short scale density perturbations evolve as predicted by the
second-order perturbation theory assumed by \cite{2012JCAP...03..004S}
in the expected regime ($\delta<1$). This is
also reflected in the optical depth fluctuations bias, which agrees
with the numerical methods in the same regime.

When we turn to the flux bias, however, the analytical method seems to
work surprisingly better than for the optical depth bias, agreeing
even in the smallest smoothing scales with the numerical methods. This
may be a reflection of the fact that, unlike the optical depth, flux
becomes more sensitive to the accurate description of less
dense regions, where in the real space FGPA framework,
$F=\mathrm{exp}(-A(1+\delta)^\alpha)$, the higher densities are
suppressed. Indeed, the analytical prediction of the flux bias in Equation 2.11,
as discussed in \cite{2012JCAP...03..004S}, is very sensitive to void regions.
These
less dense, void regions, are exactly where linear and second order
perturbation theory correctly describe the mapping between densities
in a typical region and an equivalent matter distribution riding a
large-scale over or under-density. Thus the analytical form of the
bias based on these assumptions will hold especially well for the flux
field. An argument supporting this idea is that Taylor-expanding the
optical depth field to a finite order does not reproduce the bias to
the same accuracy and hence it is not the predictions of the 2- or
3-point functions that make the 2nd order perturbation theory work but
instead the sucess in describing the matter density PDF change in the
relevant regime.

When we introduce redshift space distortions, the predictions for the density bias
parameter becomes considerably less accurate. By attempting increasingly
realistic ways of introducing the redshift-space distortions we note
that this is associated with non-linear mapping between real and
redshift-space even for linear velocities. Adding thermal broadening
further degrades any agreement. 

For the velocity gradient bias we show that the expression given in
\cite{2012JCAP...03..004S} is exact in the limit of no thermal
broadening. Our numerical results support this conclusion, effectively
by construction. However, when thermal broadening is taken into
account the predictions can be off considerably amounts. This
indicates that an appropriate combination of measurements of the small
scale flux PDF and the large scale bias parameters might provide a
path towards robust measurements of the thermal state of the IGM.

Finally, we measure bias parameters from simulations using two
different methods: a peak-background split method and direct
mode-by-mode cross power spectrum method. While we find that they in
general agree very well, we have found evidence that our 40 Mpc/h boxes
still suffer from finite box effects and that ``one loop'' bias
parameters and noise contributions might be important. It is crucial
that a physics-based parameterization of the \lyaf\ bias parameters is
developed, especially to allow for a robust measurement from
finite-sized boxes. We find that bigger $A$ values might put more emphasis 
on the more extreme/void regions of the density box, skewing the bias measurements
towards these regions which are not well represented by the size of the box. At the same time,
the fundamental $k$ mode for the mode-by-mode fit is on the verge of being non-linear. Therefore further studies
will need to include bigger box size simulations to fully understand this discrepancy.

We also note that precise measurements using multiple boxes in the PBS
formalism is difficult to do precisely: even if one has outputs
at the precise redshifts (something we do not have), there are errors
associated with external non-gravitational inputs such as reionization
which is parameterized by redshift and hence occurs at slightly wrong
times in offset boxes. For the future, it is therefore probably easier to inject a
single mode with large amplitude into a single box and measure bias
values off that particular mode. 

We conclude that while analytical methods can provide good insight into
the inner workings of the biasing of the \lya forest, they ultimately
fail to provide sufficient accuracy for precision cosmology. This is
not due to their description of non-linear density field, since the
match is excellent for real-space flux, but rather the messiness of
small-scale velocity effects.

\section*{Acknowledgements}

We would like to thank Nishikanta Khandai for providing the Gadget-3 hydrodynamic simulations, 
and Uro\v{s} Seljak, Patrick McDonald, and Zarija Luki\'c for useful discussions.

\bibliographystyle{JHEP}
\bibliography{cosmo,cosmo_preprints}

\appendix
\section{Parameters of peak-background perturbed simulations }
\label{app:pb}

Here we derive the parameters needed to evolve an over- or under-dense simulation, representing over- or under-dense regions of the fiducial cosmology. The main requirement is such that $\rho'(t) = \rho(t) (1+\delta_l(t))$, where $\delta_l$
represents the matter overdensity with respect to the background
universe (which can be thought of as the long wavelength mode
overdensity in this region of the universe). This $\delta_l$ evolves linearly with the growth factor such that $\delta_l (t) = D(t) \delta_0$ and $d\delta_l(t)/dt = H(t) f(t) \delta_l(t)$, with $D(t)$ is the linear growth factor, $\delta_0$ is the value of $\delta_l$ at $z=0$, and $f(t)$ is the derivative of the logarithmic growth factor.

Since $\rho(a) \propto a^{-3}$ and $\mathrm{dln}\rho/\mathrm{d}t = \mathrm{dln}\rho/\mathrm{d}a \dot{a} = -3 H(t)$, one can derive the hubble parameter for the perturbed simulation to first order in the long overdense mode:

\begin{align}
\frac{\mathrm{dln}\rho'(t)}{\mathrm{d}t} &= \frac{\mathrm{d}}{\mathrm{d}t} \left( \mathrm{ln} \rho(t) + \mathrm{ln}\left(1+\delta_l(t)\right)\right) \nonumber
\\ -3H'(t) & \approx -3 H(t) +H(t) f(t) \delta_l(t) \nonumber
\\ H'(t) &\approx H(t) \left(1-\frac{1}{3} f(t) \delta_l(t)\right).
\end{align}

Using this result we can derive the value of $\Omega_m$ for the perturbed simulation, again to first order in the perturbed wavelength:

\begin{align}
\rho'(t) &= H'(t) \Omega_m'(t) = H^2(t) \Omega_m \left(1+\delta_l(t)\right) \nonumber
\\ \Omega_m' &\approx \Omega_m \left[1+\left(1+\frac{2}{3} f(t)\right)\delta_l(t)\right].
\end{align}

In the same way, we derive the value of the perturbed $\Omega_\Lambda$, where we assume the dark energy density does not change:

\begin{align}
\rho'_\Lambda(t) &= \rho_\Lambda(t) = H^2(t) \Omega_\Lambda \nonumber
\\ \Omega_\Lambda'(t) &\approx \left[1 + \frac{2}{3} f(t) \delta_l(t) \right] \Omega_\Lambda(t).
\end{align}

Since the perturbed simulations will now have different cosmologies, the evolution of time with redshift is different as well. In order to use these over or underdense simulations at the same time (not redshift) as the fiducial simulations, we derive at what redshift outputs the perturbed simulations will have the same time as the unperturbed simulations. Using the relation $\rho(t) = \rho_0 (1+z)^3$ where $\rho_0$ is the density at $z=0$ of the fiducial simulation, we require that all times the ratio $\frac{\rho'(t)}{\rho(t)} = 1+\delta_l(t) = \frac{\rho'_0(1+z')^3}{\rho_0(1+z)^3}$ is satisfied. Rearranging this ratio we then arrive at:

\begin{align}
(1+z')^3 &= (1+z)^3 \frac{\rho_0}{\rho'_0}\left(1+\delta_l(t)\right) \nonumber
\\ & \approx (1+z)^3 \frac{1+\delta(t)}{1+\delta_0}, \nonumber
\end{align}
giving us:

\begin{equation}
\label{eq:pbredshift}
1+z' \approx (1+z)\left[1+\frac{1}{3}\left(\delta_l(t)-\delta_0\right)\right].
\end{equation}

Lastly, since the density changes, so will the value of $\sigma_8$. Since this value also scales with the growth factor, we can express it in terms of the value at $z=0$:

\begin{equation}
\sigma_8 (z) = \sigma_8 D(z)/D(0). \nonumber
\end{equation}

Then we can use the fact that the ratio of this value of the unperturbed to the perturbed simulation approaches 1 at infinite redshift
\begin{equation}
  \lim_{a \rightarrow 0} \frac{\sigma_8 D(z)/D(0)}{\sigma_8' D'(z')/D'(0)}=1,
\end{equation}
thus arriving at:

\begin{align}
\sigma'_8 &= \sigma_8 \frac{D'(0)}{D(0)}\frac{D(z)}{D'(z')} \nonumber
\\& =  \sigma_8 \frac{g'(z=0)}{g(z=0)}\frac{1+z}{1+z'} \vert_{z=0} \frac{g(z)}{g'(z')} \frac{1+z'}{1+z}\vert_{z\rightarrow \infty} \nonumber
\\& \approx \sigma_8 \frac{g'(z=0)}{g(z=0)} \frac{1+z'}{1+z}\vert_{z\rightarrow \infty} \nonumber
\\& \approx \sigma_8 \frac{g'(z=0)}{g(z=0)} \left[ 1 +\frac{1}{3} \left(\delta(z\rightarrow \infty)-\delta_0\right) \right]\nonumber
\\& \approx  \sigma_8 \frac{g'(z=0)}{g(z=0)}\left[ 1 -\frac{1}{3} \delta_0\right],
\end{align}
where we used the fact that $D(a)=g(a) a$ and $g(a)\rightarrow1$ as $a \rightarrow 0$. Here we also pick our equality point such that $z=z'=0$ at the same time $t_0$. To derive the parameters in 
Table \ref{tab:pbsims}, we read off these derived perturbed parameters at $z=z'=0$, where we pick $\delta_l (z=2.5) = 0.015$, which is then scaled to arrive at the corresponding value of $\delta_0$ at $z=0$ for the formulae above.

\end{document}